\documentclass[journal=polymersau,manuscript=article]{achemso}

\usepackage{chemformula} 
\usepackage[T1]{fontenc} 
\usepackage{bm}
\usepackage{xcolor}
\usepackage{amssymb}
\usepackage{amsmath}

\newcommand\dmi[1]{{\color{black}#1}}

\DeclareMathOperator*{\argmax}{arg\,max}
\author{Joseph L. Sleiman}
\affiliation[Edinburgh]{School of Physics and Astronomy, University of Edinburgh, Peter Guthrie Tait Road, Edinburgh, EH9 3FD, UK}
\author{Robin H. Burton}
\affiliation[Edinburgh]{School of Physics and Astronomy, University of Edinburgh, Peter Guthrie Tait Road, Edinburgh, EH9 3FD, UK}
\author{Michele Caraglio}
\affiliation[Austria]{Institut f\"{u}r Theoretische Physik, Universit\"{a}t Innsbruck, Technikerstra{\ss}e 21A, A-6020, Innsbruck, Austria}
\author{Yair Augusto Gutierrez Fosado}
\affiliation[Edinburgh]{School of Physics and Astronomy, University of Edinburgh, Peter Guthrie Tait Road, Edinburgh, EH9 3FD, UK}
\author{Davide Michieletto}
\affiliation[Edinburgh]{School of Physics and Astronomy, University of Edinburgh, Peter Guthrie Tait Road, Edinburgh, EH9 3FD, UK}
\alsoaffiliation[MRC]{MRC Human Genetics Unit, Institute of Genetics and Cancer, University of Edinburgh, Edinburgh EH4 2XU, UK}
\email{davide.michieletto@ed.ac.uk}

\title{Geometric Predictors of Knotted and Linked Arcs} 

\abbreviations{IR,NMR,UV}
\keywords{Polymers; Entanglements; Topology; Knots; Links; Writhe;}

\begin{document}

\begin{tocentry}
\includegraphics{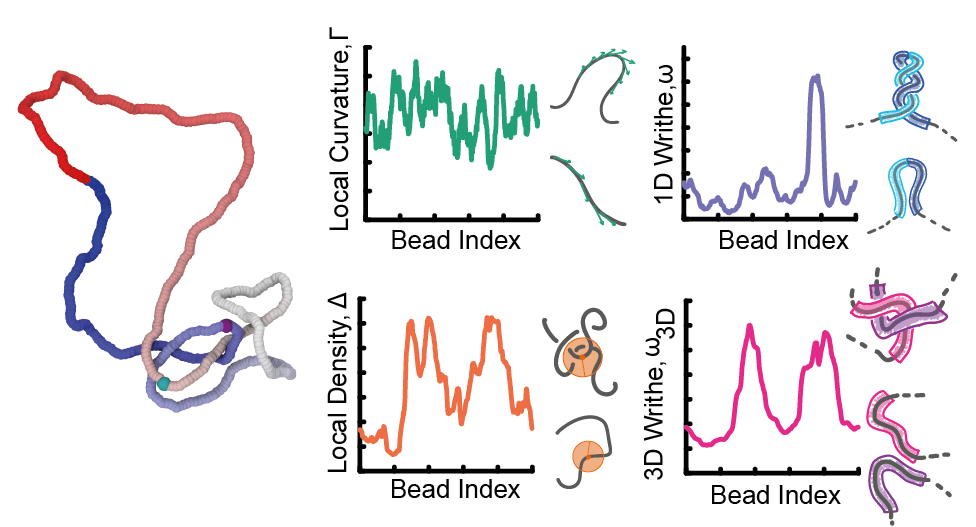}
\end{tocentry}


\begin{abstract}
   Inspired by how certain proteins ``sense'' knots and entanglements in DNA molecules, here we ask if there exist local geometric features that may be used as a read-out of the \dmi{underlying topology of generic polymers}. We perform molecular simulations of knotted and linked \dmi{semiflexbile polymers} and study four geometric measures to predict topological entanglements: local curvature, local density, local 1D writhe and non-local 3D writhe. We discover that local curvature is a poor predictor of entanglements. In contrast, segments with maximum local density or writhe correlate as much as 90\% of the time with the shortest knotted and linked arcs. We find that this accuracy is preserved across different knot types and also under significant spherical confinement, which is known to delocalise essential crossings in knotted polymers. We further discover that non-local 3D writhe is the best geometric read-out of knot location. Finally, we discuss how these geometric features may be used to computationally analyse entanglements in generic polymer melts and gels. 
\end{abstract}


\section*{Introduction} 

Topological entanglements are ubiquitous, and an essential feature of everyday materials and complex fluids, endowing them with viscous and elastic properties. Entanglements are often poorly defined and their unambiguous identification and quantification remains elusive~\cite{Panagiotou2013,Igram2016}. For example, a knot is a well defined mathematical entity when tied on a closed curve, but there are many examples in physics and biology, e.g. proteins and chromatin, where knots are tied on open curves, rendering such ``physical'' knots much more difficult to define rigorously and unambiguously~\cite{Tubiana2011,Goundaroulis2017,Goundaroulis2017a,Goundaroulis2019,Panagiotou2020}. 
More broadly, a long-standing goal in polymer physics and the broader soft matter communities is to understand and control the topology of certain systems from the geometry of (often entangled) 1D curves. This goal encompasses many fields, from liquid crystals~\cite{Machon2013}, optics~\cite{OHolleran2008,Dennis2010}, fluids~\cite{Laing2015}, DNA~\cite{Smrek2013,Siebert2017,Goundaroulis2019,Michieletto2017,Marenduzzo2009,Klotz2020kdna,Klotz2020knots,Polson2021}, proteins~\cite{Baiesi2017,Dabrowski-Tumanski2017}, polymers~\cite{Tubiana2011,Marenduzzo2010,Wu2017,Rauscher2018}, soap films~\cite{Goldstein2010b,Machon2016}, and soft matter in general~\cite{Kamien2002,Dennis2005}. \dmi{At the same time, the unambiguous characterisation of entanglements in these systems are often elusive, in turn begging for better strategies to quantify entanglements in generic soft matter systems. }

\dmi{A striking example of the inherent difficulty in defining entanglements is seen in polymer melts, whereby the close contact of two chains does not necessarily indicate that chains are constraining each other's motion. Instead, so-called ``primitive''~\cite{Everaers2004} and ``isoconfigurational''~\cite{Bisbee2011} mean path techniques are far better placed to separate relevant entanglements from irrelevant ones. Yet, even these sophisticated techniques often struggle when polymers display non-trivial topology, e.g. rings~\cite{McLeish2002,Halverson2011statics,Rosa2013}. Ring polymers are in fact not amenable for standard primitive path analysis as they do not entangle in the traditional sense as linear polymers do~\cite{Halverson2011dynamics,Halverson2011statics}; e.g. no ``tube'' can be defined around their contour and they do not ``reptate''~\cite{Halverson2011dynamics}. Rings display architecture-specific topological constraints called threadings~\cite{michieletto2014threading,Gomez2020} which display the puzzling property to reduce self-similarly over time~\cite{Ge2016,michieletto2020dynamical}. Developing a method to robustly and unambiguously quantifying entanglements in melts of ring polymers is still an open challenge in the field of polymer physics~\cite{Smrek2016,landuzzi2020persistence}. }

\dmi{In parallel to these open questions, it is clear that the geometric design of systems with specific entanglements in their microstructure could in principle allow for the control of mesoscopic material properties~\cite{Evans2014,Igram2016,Oster2021}. The realisation of woven structures can now be achieved at both micro- and meso-scales using synthetic chemistry~\cite{August2020} or 3D printing~\cite{Oster2021}. To bypass a virtually endless trial-and-error approach, it is therefore important to be able to select the entanglement motifs to embroider in the structures in such a way that they display the desired mechanical properties~\cite{August2020}. Interestingly, this problem is not too dissimilar to that of knitting socks: using solely two types of stitches (``knit'' and ``purl''), it is possible to create many distinct motifs, and socks with distinctive elastic properties~\cite{Matsumoto2018,Markande2020}.
}


\begin{figure*}[t!]
  \centering
  \includegraphics[width=0.95\textwidth]{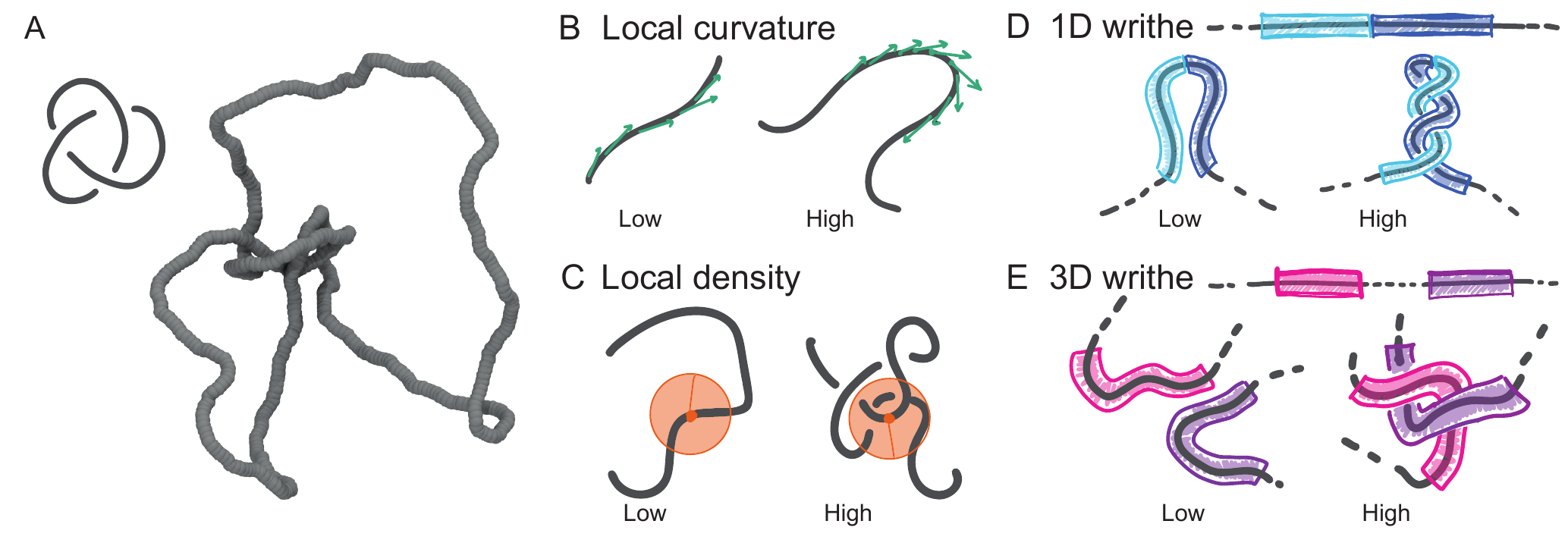}
\caption{\textbf{A.} Snapshot of a trefoil knot during a molecular dynamics simulation. \textbf{B-E.} Illustration of the four geometric descriptors considered in this work. \textbf{B.} Local curvature (Eq.~\eqref{eqn:lc}), \textbf{C.} local density (Eq.~\eqref{eq:ld}), \textbf{D.} 1D writhe (Eq.~\eqref{eq:1dwrithe}), and \textbf{E.} 3D writhe (Eq.~\eqref{eq:3dwrithe}).}
\label{fig:descriptors}
\end{figure*}

\dmi{Another example in which topological entanglements are abundant is in molecular biology and genome organisation. Two meters of genome is packed in a 10 $\mu$m nucleus in human cells. This extreme level of packaging leads to knotting and entanglement which are resolved by Topoisomerase (Topo2)  -- a protein that is about $50$ nm in size -- which can identify topological knots from pure geometric entanglements in DNA molecules that are more than a thousand times bigger~\cite{Wang1985}. By a still poorly understood ``sensing'' mechanism~\cite{Martinez-Garcia2014,Michieletto2022nar}, Topo2 is able to reduce the topological complexity of DNA {\it in vivo}~\cite{Martinez-Garcia2014,Piskadlo2017,Valdes2018} and {\it in vitro}~\cite{Rybenkov1997} without introducing more complex knots.}


Inspired by Topo2's topological sensing - which is necessarily local, unable to account for the global topology of knotted DNA - here we investigate the possibility that there exist some geometric descriptors that correlate with the underlying topology of \dmi{generic closed curves involved, for instance, in woven structures or polymer melts}. To this end, we perform molecular dynamics (MD) simulations of knotted and linked semiflexible polymers in equilibrium and study the correlation between the position of the shortest knotted and linked arcs with that of four geometric descriptors: (i) regions of maximum local curvature, (ii) regions of maximum local density, (iii) regions with maximum local 1D writhe and (iv) regions with maximum non-local 3D writhe. 
\dmi{We note that while Topo2 works on a very specific polymer - the DNA double-helix - here we are interested in exploring the relationship between local geometry and global topology on generic polymers, with the hope that our results may be helpful for better understanding entanglements in generic entangled polymer systems. }

In this paper we discover that regions of maximum local density strongly correlate with knotted and linked arcs and outperform regions of maximum curvature. Surprisingly, we also find that this effect persists under strong confinement, where the knotted polymer is confined within a sphere smaller than its size in equilibrium. Finally, we show that 3D writhe is the best geometrical descriptors to recognise knotted arcs, and it performs consistently better than other geometric predictors. We \dmi{conjecture} that these local geometric descriptors could be employed to compute topological entanglements in more complex systems such as \dmi{polymer melts, networks, tangles and weavings}.

\section*{Methods}

\begin{figure*}[t!]
  \centering
  \includegraphics[width=0.9\textwidth]{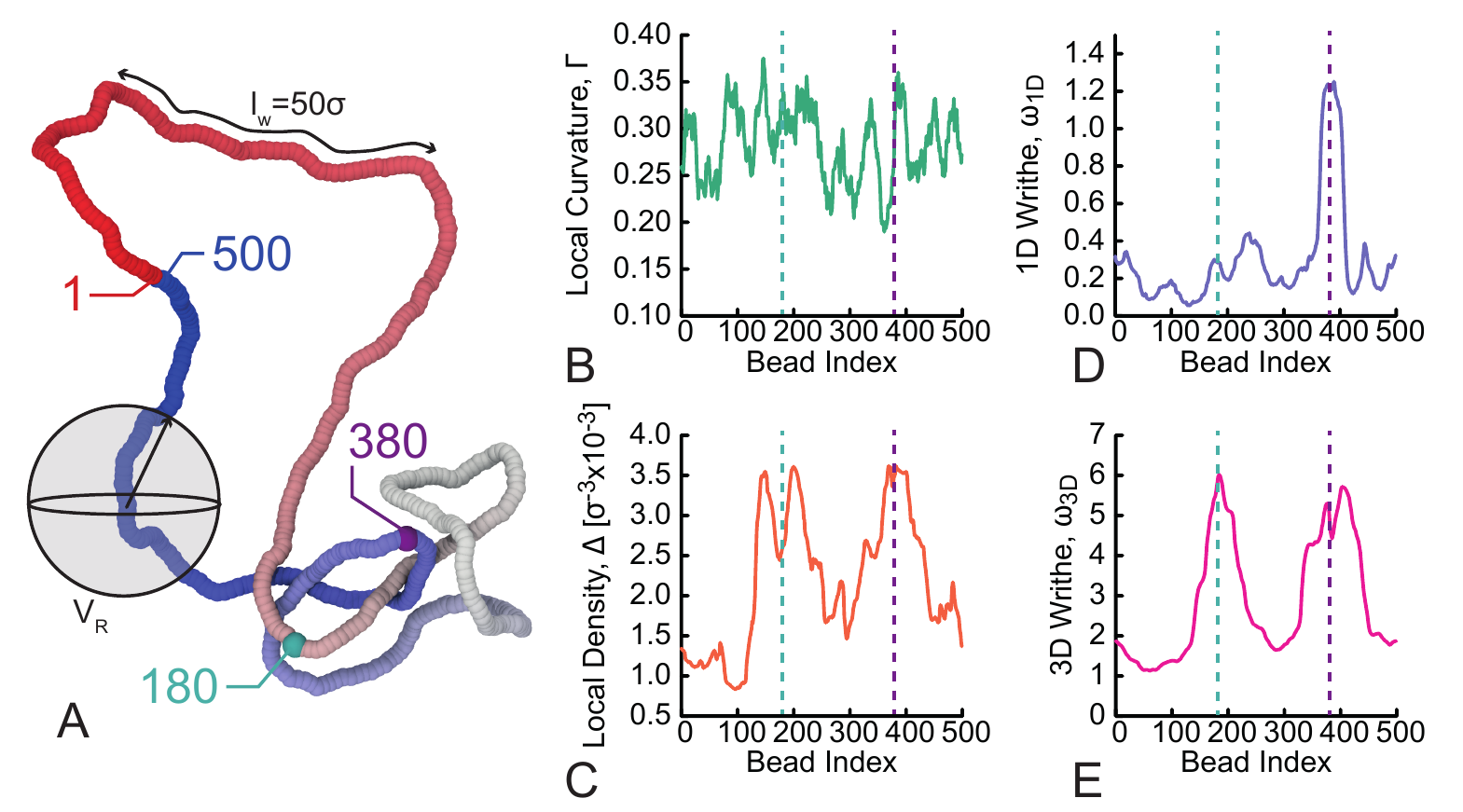}
\caption{\textbf{A.} Snapshot of a simulated trefoil knot conformation, color coded in terms of the bead index. \dmi{The window $l_w=50\sigma$ used for the 1D and 3D writhe and the sphere volume $V_R$ of radius $R=30\sigma$ used for the local density are also shown}. \textbf{B-E.} Curves obtained via the calculation of the geometric descriptors defined in the text and computed on the configuration in \textbf{A}: \textbf{B.} local curvature (Eq.~\eqref{eqn:lc}), \textbf{C.} local density (Eq.~\eqref{eq:ld}), \textbf{D.} 1D writhe (Eq.~\eqref{eq:1dwrithe}) and \textbf{E.} 3D writhe (Eq.~\eqref{eq:3dwrithe}). The beads (180 and 380) corresponding to peaks in the local density and 3D writhe are highlighted in \textbf{A-E.}}
\label{fig:geom_example}
\end{figure*}

\subsection*{Simulation details}
We model knotted and linked curves as semiflexbile coarse-grained bead-spring polymers with $N=500$ beads of size $\sigma$. The beads interact with each other via a purely repulsive Lennard Jones potential
\begin{equation}\label{eq:LJ}
U_{\rm LJ}(r) = \left\{
\begin{array}{lr}
4 \epsilon \left[ \left(\dfrac{\sigma}{r}\right)^{12} - \left(\dfrac{\sigma}{r}\right)^6 + \dfrac14 \right] & \, r \le r_c \\
0 & \, r > r_c
\end{array} \right. \, ,
\end{equation}
where $r$ denotes the separation between the beads and the cutoff $r_c=2^{1/6}\sigma$ is chosen so that only the repulsive part of the potential is used. Nearest-neighbour monomers along the contour of the chains are connected by  finitely extensible nonlinear elastic (FENE) springs as
\begin{equation}\label{eq:Ufene}
U_{\rm FENE}(r) = \left\{
\begin{array}{lcl}
-0.5kR_0^2 \ln\left(1-(r / R_0)^2\right) & \ r\le R_0 \\ \infty & \
r> R_0 &
\end{array} \right. \, ,
\end{equation}
where $k = 30\epsilon/\sigma^2$ is the spring constant and $R_{0}=1.5\sigma$ is the maximum extension of the elastic FENE bond. This choice of potentials and parameters is essential to preclude thermally-driven strand crossings and therefore ensures that the global topology is preserved at all times~\cite{Kremer1990}.  Finally, we add bending rigidity via a Kratky-Porod potential, $U_{\rm bend}(\theta) = k_\theta \left(1 - \cos \theta \strut\right)$,  where $\theta$ is the angle formed between consecutive bonds and $k_\theta=20 k_BT$ is the bending constant. We choose this value to mimic that of DNA, as for $\sigma=2.5$ nm the persistence length would be matched to $l_p = 50$ nm, as known for DNA~\cite{Calladine1997}. Each bead's motion is then evolved via a Langevin equation, i.e. by adding to the Newtonian equations of motion a friction and stochastic term related by the fluctuation-dissipation relation, where the amplitude of the stochastic delta-correlated force is given by $\sqrt{2k_BT/\gamma}$, and $\gamma$ is the friction coefficient. The numerical evolution of the system is conducted using a velocity-verlet scheme with $dt=0.01 \tau_{LJ} = 0.01 \sigma \sqrt{m/\epsilon}$ in LAMMPS~\cite{Plimpton1995}. In order to simulate knotted chains, we initialise the chain of beads by using the well-known parameterisation for torus knots:  $(x,y,z)(t) = ( R(\cos{q t} + r) \cos{p t}, R(\cos{q t} + r) \sin(p t), - R \sin(q t))$ where $p$ and $q$ are  co-prime integers,  $R$ and $r$ are two constants, and $t\in \left(0, 2 \pi\right)$. 

In our paper, we want to compute the likelihood that some geometric features (to be defined below) yield accurate predictions of where the shortest knotted or linked segments are. To do this we typically consider 1000 configurations taken by dumping the coordinates of the beads every $10^4$ LAMMPS steps (or $10^2 \tau_{LJ}$). 
From each simulation we obtain the fraction of instances in which our predictors (described below) correctly identify the knotted or linked arc. We then run 64 independent replicas (starting from different initial conformations) and typically plot the distribution of this fraction in the form of boxplots (see below for details).

\section{Results}

\subsection{Geometric Descriptors}
As mentioned above, we consider four geometric descriptors that allow us to map polymer beads to a scalar quantity (see Fig.~\ref{fig:descriptors} for a visual representation). They are (i) local polymer curvature (see Fig.~\ref{fig:descriptors}B) 
\begin{equation}
\label{eqn:lc}
    \Gamma(i) = \dfrac{1}{n} \sum_{j=i-n/2}^{i+n/2}\arccos\left({\frac{\bm{t}_{j-1,j} \cdot \bm{t}_{j,j+1}}{\lvert\bm{t}_{j-1,j}\rvert\lvert\bm{t}_{j,j+1}\rvert}}\right)
\end{equation}
where $\bm{t}_{j,j+1} \equiv \bm{r}_{j+1}-\bm{r}_{j}$ is the tangent vector at bead $j$ and $n=20$ an averaging window. (ii) Local bead density (see Fig.~\ref{fig:descriptors}C)
\begin{equation}
\Delta(i) = \dfrac{1}{V_R}\sum_{j \neq i}^N \Theta(R - |\bm{r}_i - \bm{r}_j|) 
 \label{eq:ld}
\end{equation}
where $\Theta(x)=1$ if $x>0$ and 0 otherwise. In this equation, $V_R = 4\pi R^3/3$ is the volume of a sphere of radius $R$, and we take $R=30 \sigma$, slightly larger than a persistence length. We have checked that other sensible choices of $R$ give similar results. (iii) Local or 1D (unsigned) writhe (see Fig.~\ref{fig:descriptors}D) as   
\begin{equation}
\omega_{1D}(i) = \dfrac{2}{4\pi} \sum_{k=i-l_w}^{i} \sum_{l=i}^{i+l_w}  \dfrac{\left| (\bm{t}_{k,k+1} \times \bm{t}_{l,l+1}) \cdot (\bm{r}_k - \bm{r}_l)\right|}{|\bm{r}_k - \bm{r}_l|^3} \, 
 \label{eq:1dwrithe}
\end{equation}
where $l_w=50 \sigma$ is the window length over which the calculation is performed. Finally (iv) non-local or 3D (unsigned) writhe (see Fig.~\ref{fig:descriptors}E) as
\begin{equation}
\omega_{3D}(i) = \dfrac{2}{4\pi}\sum_{k=i-\frac{l_w}{2}}^{i+\frac{l_w}{2}} \sum_{l=0}^{N} \dfrac{\left| (\bm{t}_{k,k+1} \times \bm{t}_{l,l+1}) \cdot (\bm{r}_k - \bm{r}_l)\right|}{|\bm{r}_k - \bm{r}_l|^3}  \, 
 \label{eq:3dwrithe}
\end{equation}
which measures the (unsigned) entanglement of a polymer length centred at bead $i$ against the rest of the polymer contour.

Eq.~\eqref{eq:1dwrithe} is the local generalisation of the well-known ``average crossing number''~\cite{Stasiak1996} and has been previously used to identify supercoiled plectonemes in simulated DNA~\cite{Klenin2000,Smrek2021supercoiling}, branches in ring polymers~\cite{Michieletto2016a} and self-entanglements in proteins~\cite{Baiesi2017}. Eq.~\eqref{eq:3dwrithe} is a generalisation of Eq.~\eqref{eq:1dwrithe} where we do not restrict the calculation of the (unsigned) writhe to occur between contiguous polymer segments. Intuitively, Eqs.~\eqref{eq:1dwrithe} and \eqref{eq:3dwrithe} effectively compute the average number of times the contiguous (for 1D) and non-contiguous (for 3D) segments of the polymer display crossings when observed from many different directions. Accordingly, we define the beads at which our descriptors attain their maximum value as $i_{X} = \argmax_i \left\{ X \right\}$, where $X = \{ \Delta(i), \Gamma(i), \omega_{1D}(i), \omega_{3D}(i)\}$. 

Examples of typical curves that we get from the calculation of these observables on simulated polymers are shown in Fig.~\ref{fig:geom_example}. The snapshot in Fig.~\ref{fig:geom_example}A has been color-coded from red to blue to identify the bead index. Beads 180 and 380 are colored green and purple to highlight the correspondence with the curves on the right of Fig.~\ref{fig:geom_example}. One can appreciate that the local curvature $\Gamma$ (Fig.~\ref{fig:geom_example}B) is rather noisy and does not seem to reflect an increase in entanglements around beads 180 and 380. On the contrary, local density $\Delta$ (Fig.~\ref{fig:geom_example}C) displays three local maxima corresponding to increased density of 3D proximal segments around beads 180 and bead 380. Strikingly, 1D writhe $\omega_{1D}$ and 3D writhe $\omega_{1D}$ (Figs.~\ref{fig:geom_example}D,E) display the most intuitive and marked trends. The 1D writhe $\omega_{1D}$ is best suited to detect self-entanglements over short distances (around $l_w$), while the 3D writhe $\omega_{3D}$ is able to detect self-entanglements over large distances. Intuitively, the peaks correspond to the location of the essential crossings of the trefoil knot.

\begin{figure*}[t!]
  \centering
  \includegraphics[width=0.9\textwidth]{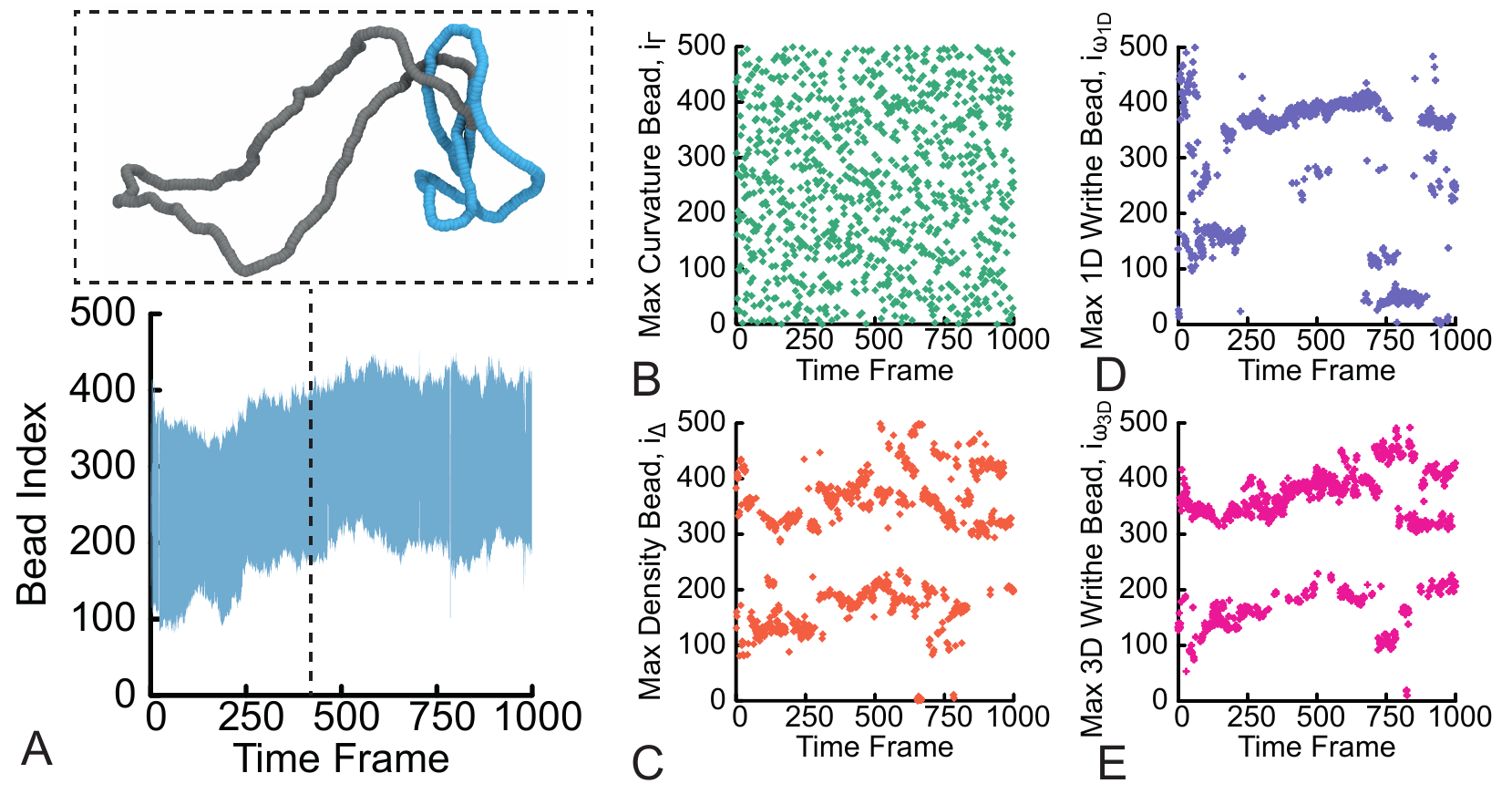}
\caption{In this figure we report kymographs (the evolution of geometric and topological observables over time). In \textbf{A}, we show in shaded blue the range of beads identified by Kymoknot that form the shortest knotted arc during one simulation of a trefoil knot. The inset shows a snapshot of the simulation, corresponding to an instantaneous conformation with the shortest knotted arc color coded in blue. In \textbf{B-E} we show the argmax value of (\textbf{B}) local curvature, (\textbf{C}) local density, (\textbf{D}) 1D writhe, and (\textbf{E}) 3D writhe at each time frame during the molecular dynamics simulation. }
\label{fig:kymo}
\end{figure*}

\subsection{Knot Localisation}
To identify knotted arcs in our simulated polymer we use Kymoknot~\cite{Tubiana2018}, a free and open-source software to identify the topology and shortest knotted arcs of closed and open polymer chains. The algorithm works by using a minimally interfering algorithm that (either in a top-down or bottom-up direction) truncates the polymer conformation, computes the convex hull of the remaining polymer segments, joins the termini outside the so-formed convex hull and then calculates the Alexander determinant of the closed conformation~\cite{Tubiana2011}. The result of Kymoknot is the interval within which the shortest knotted arc is located. For a polymer conformation that evolves in time, we can visualise the output of Kymoknot in a so-called kymograph. The blue shaded region in Fig.~\ref{fig:kymo}A represents the shortest knotted arc within the simulated polymer as it fluctuates in time. For clarity, we also show a representative snapshot of the polymer at a given time frame where we have color-coded the shortest knotted arc in blue. We then directly use the Kymoknot output to count how frequently the $i_X$'s computed using the geometric descriptors defined above fall within the shortest knotted interval. We call this quantity the ``colocalisation score'', $\rho_X$.  

The key point of this work is that Kymoknot recognises the shortest knotted arc by computing a global topological invariant (the Alexander determinant) of a suitably closed open curve. On the contrary, the quantities defined in Eqs.~\eqref{eqn:lc}-\eqref{eq:3dwrithe} are purely geometric and have no knowledge of the global topology of the chain. Additionally, 3 of them $\Delta$, $\Gamma$, and $\omega_{1D}$ are purely \emph{local} features that can be extracted from a short polymer segment, measuring the surrounding segments in close 1D or 3D proximity.

\begin{figure}[t!]
  \centering
 \includegraphics[width=0.45\textwidth]{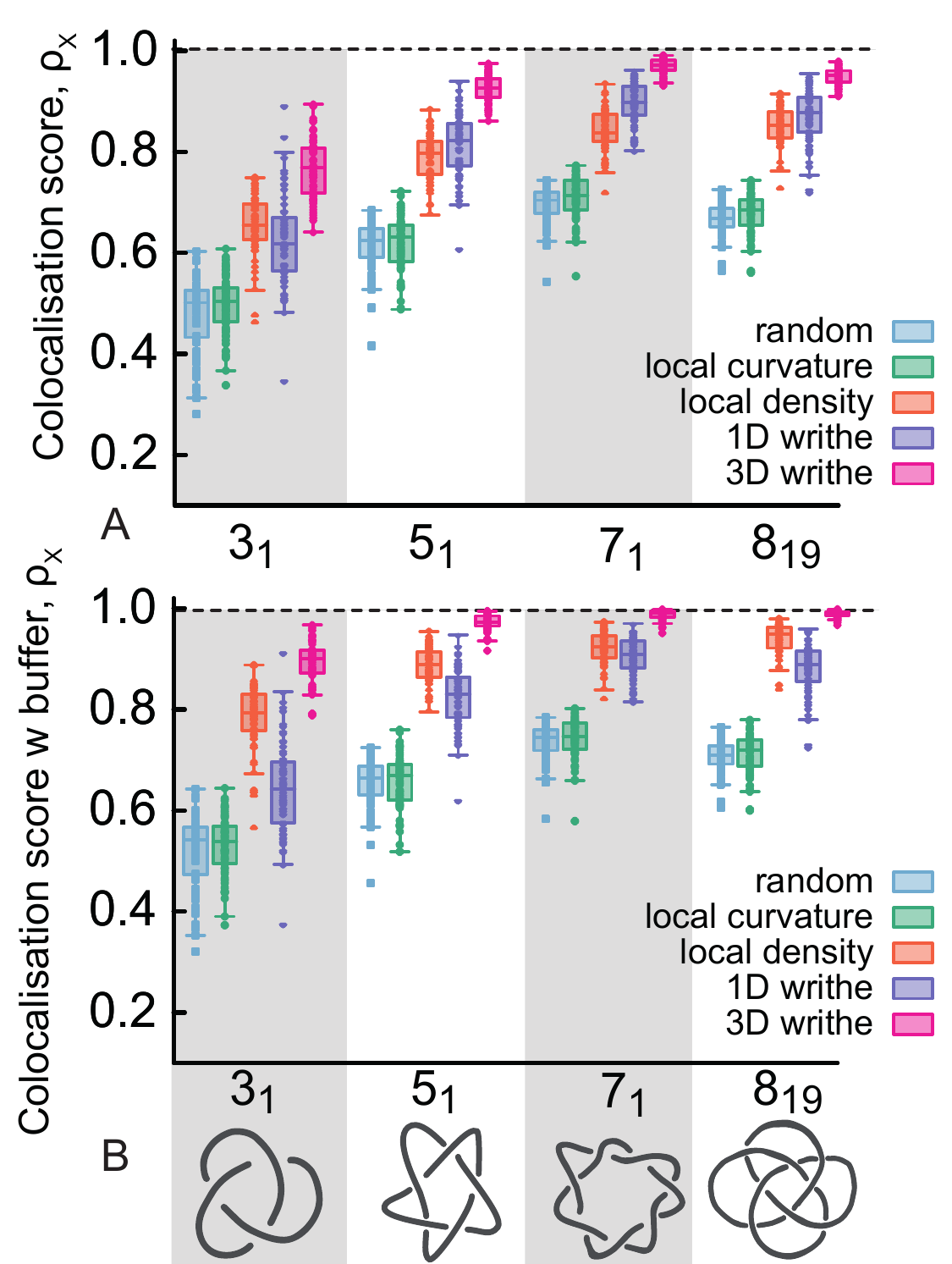}
\caption{\textbf{A.} Boxplots showing the colocalisation score of four different knot types using the four geometric descriptors (plus a random control) over 64 replicas. Each point in the boxplot represents the colocalisation score (i.e. how many times the geometric predictor is contained within the Kymoknot-detected arc) computed over 1000 conformations in each replica. \textbf{B.} Same as \textbf{A}, but accounting for a ``buffer'' of 10 beads on either side of the boundaries detected by Kymoknot.}
\label{fig:coloc_knots}
\end{figure}

\subsection{Localisation of knotted arcs by geometric descriptors}
Having described the topological and geometrical observables used in this work to identify knotted and linked arcs, we now aim to address how well the geometric descriptors can predict the location of knots along polymers. To achieve this, we first visually compare the result from Kymoknot (Fig.~\ref{fig:kymo}A) to the ones obtained via the $i_X$'s of the geometric descriptors (Fig.~\ref{fig:kymo}B-E). We first notice that the maximum of the local curvature $\Gamma$ appears to be noisy and randomly scattered along the contour. \dmi{This is also the case if we do not perform the window averaging of the local curvature or if we pick beads separated by a number of beads}. On the contrary, the maximum of local density, 1D writhe and 3D writhe appear to locate near the boundaries of the shortest knotted arc identified by Kymoknot (Fig.~\ref{fig:kymo}A). We \dmi{hypothesise} that this finding may be related to the concept of essential crossings~\cite{Suma2017,Coronel2018} and that our geometric predictors may thus be able to identify some of the essential crossings in the knotted chain. 


\dmi{To more precisely quantify how well our predictors can identify the location of the shortest knotted arc, we compute the ``colocalisation score'', $\rho_X$. [We recall that this was defined as the number of times that the geometrically predicted $i_X$ falls within the shortest knotted interval detected by Kymoknot.]} Fig.~\ref{fig:coloc_knots}A shows that for an unconfined, dilute polymer, $\rho_\Gamma$ is similar to one obtained by a random choice of bead, i.e. for a trefoil $\rho_{rand} \simeq 50\%$. Notice that a computed $\rho_{rand} \simeq 0.5$ means that, for our choice of parameters, the shortest knotted arc occupies about half of the polymer contour; this is due to the large polymer stiffness chosen to match that of DNA and the net effect is that the knot tends to delocalise~\cite{Tubiana2011}. Interestingly, we observe a much larger colocalisation score for the other geometric descriptors. More specifically, the local density descriptor $i_\Delta$ colocalises with the knotted arc roughly $\rho_\Delta = 70\%$ of the times for a trefoil and more than 80\% for the other knot types (Fig.~\ref{fig:coloc_knots}A). Additionally, we find that the 3D writhe is the most accurate predictor, with $\rho_{\omega_{3D}} \simeq 80\%$ for the trefoil and $\rho_{\omega_{3D}} > 90\%$ for the more complex knots.

Interestingly, if we account for a ``buffer'', i.e. an additional 10 beads on either side of the knot boundaries identified by Kymoknot, we find a further increase in accuracy (see Fig.~\ref{fig:coloc_knots}B), with $i_\Delta$ reaching more than 80\% for all knot types and 3D writhe more than 90\% for all knot types, getting close to 100\% for $5_1$, $7_1$ and $8_{19}$. While local density improves its predictive power when including the buffer, the 1D writhe does not. Perhaps the most interesting observation from Fig.~\ref{fig:coloc_knots} is that even if more complex knots delocalise and take up a larger fraction of the polymer contour (see the random value increasing up to $\simeq$ 75\%), our geometric descriptors are still significantly more accurate than simply a random choice.

\subsection{Localisation of knotted arcs under spherical confinement}

\begin{figure}[t!]
  \centering
 \includegraphics[width=0.45\textwidth]{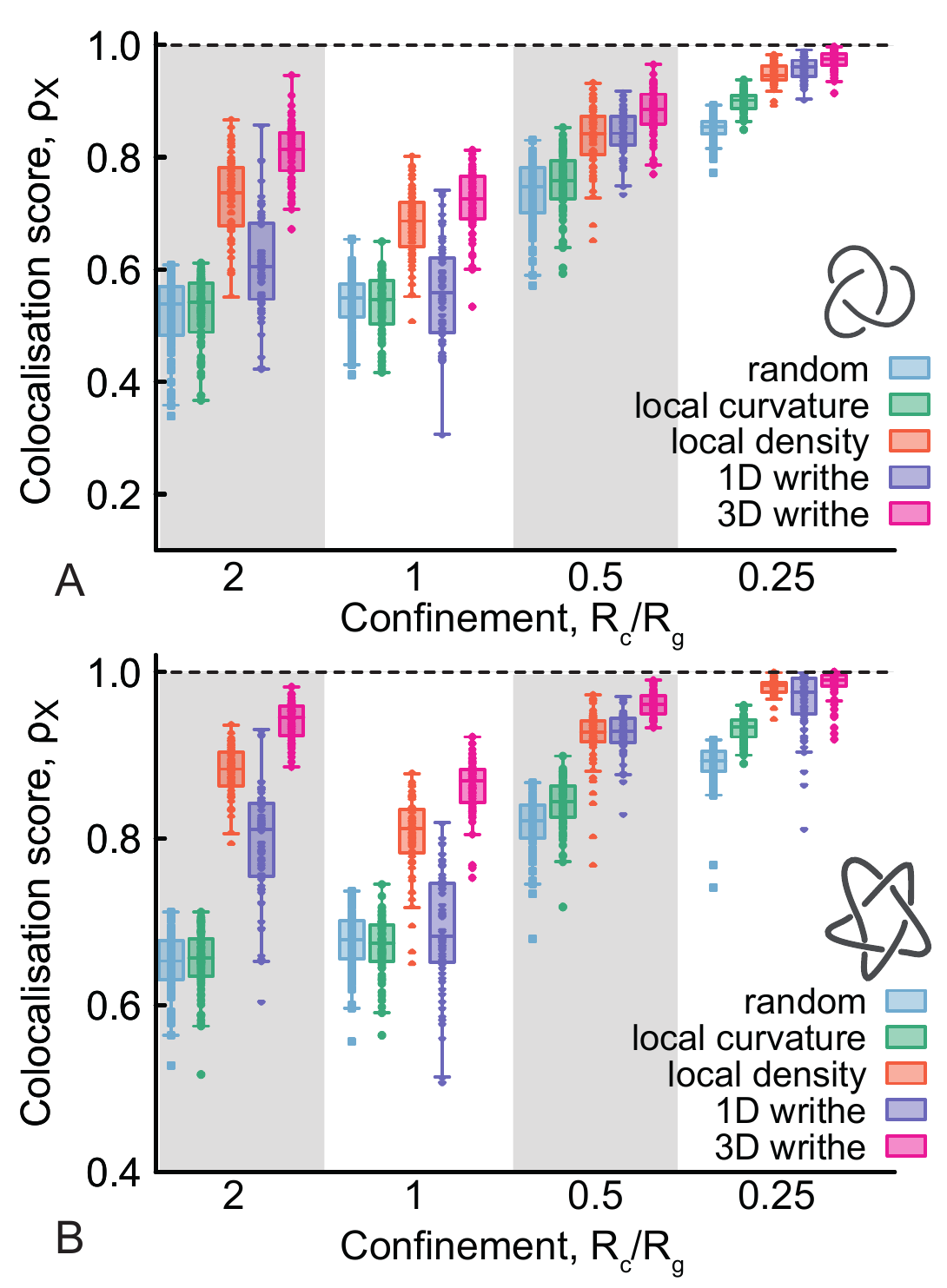}
\caption{\textbf{A-B.} Boxplots showing the colocalisation score for a trefoil (\textbf{A}) and pentafoil (\textbf{B}) as a function of knot confinement, measured as $R_c/R_g$ where $R_g$ is the radius of gyration of the polymer in equilibrium. As before, we compute the score over 1000 conformations and make the boxplot using one value for each of the 64 independent replicas.}
\label{fig:coloc_conf}
\end{figure}

Arguably, while the semiflexible nature of our chains renders knots rather delocalised over the contour, considering chains that are more flexible would induce knot localisation~\cite{Grosberg2007,Tubiana2013}, which is expected to facilitate their recognition by our geometric methods. Localised knots are defined such that their subtended arc scales sublinearly with the length of the polymers, i.e. $l_k \sim N^{\alpha}$ with $\alpha<1$. It was previously shown that knots in flexible chains display $\alpha \simeq 0.75$~\cite{Tubiana2011}. 
On the other hand, under spherical confinement knots are extremely delocalised and display $\alpha \simeq 1$~\cite{Tubiana2011}. Thus, we ask whether our geometric predictors (and in particular the local density $\Delta$) remain good predictors of knot location under spherical confinement. To study this regime, we enclose polymers in spherical shells with harmonic repulsive interactions with all the beads. The radius of the shell $R_c$ is slowly reduced until the desired confinement $R_c/R_g$ (with $R_g$ being the equilibrium radius of gyration of the polymer in dilute conditions) is attained. The polymer is then allowed to equilibrate. Finally, we measure the curves $\Gamma(i)$, $\Delta(i)$, $\omega_{1D}(i)$ and $\omega_{3D}(i)$ as before and, in turn, the colocalisation score, $\rho_X$ (Fig.~\ref{fig:coloc_conf}). The only change is that we now use $R = R_c/8$ to compute $\Delta(i)$. This is needed because under confinement the radius of gyration becomes smaller than the original value $R=30\sigma$ we set earlier for the dilute case. We have repeated this calculation for other sensible choices of $R$ and they produce qualitatively similar results. Interestingly, we observe that $\Delta$ still outperforms a random process even at values of confinement strength $R_c/R_g= 0.25$ for both the trefoil and pentafoil knots (see Fig.~\ref{fig:coloc_conf}). It is rather striking that $i_\Delta$ colocalises with the knotted arc more than $\rho_\Delta > 95\%$ of the time, meaning that even under these extreme conditions of self-density, the presence of a knot can be identified via purely geometric features.  

Finally, we note that the accuracy trend displays a non-monotonic behaviour as a function of confinement strength. In particular, we note a curious dip in accuracy for $R_c/R_g=1$. It would be interesting in the future to explore in detail the physical origin of this behaviour.

\begin{figure*}[t!]
  \centering
 \includegraphics[width=1\textwidth]{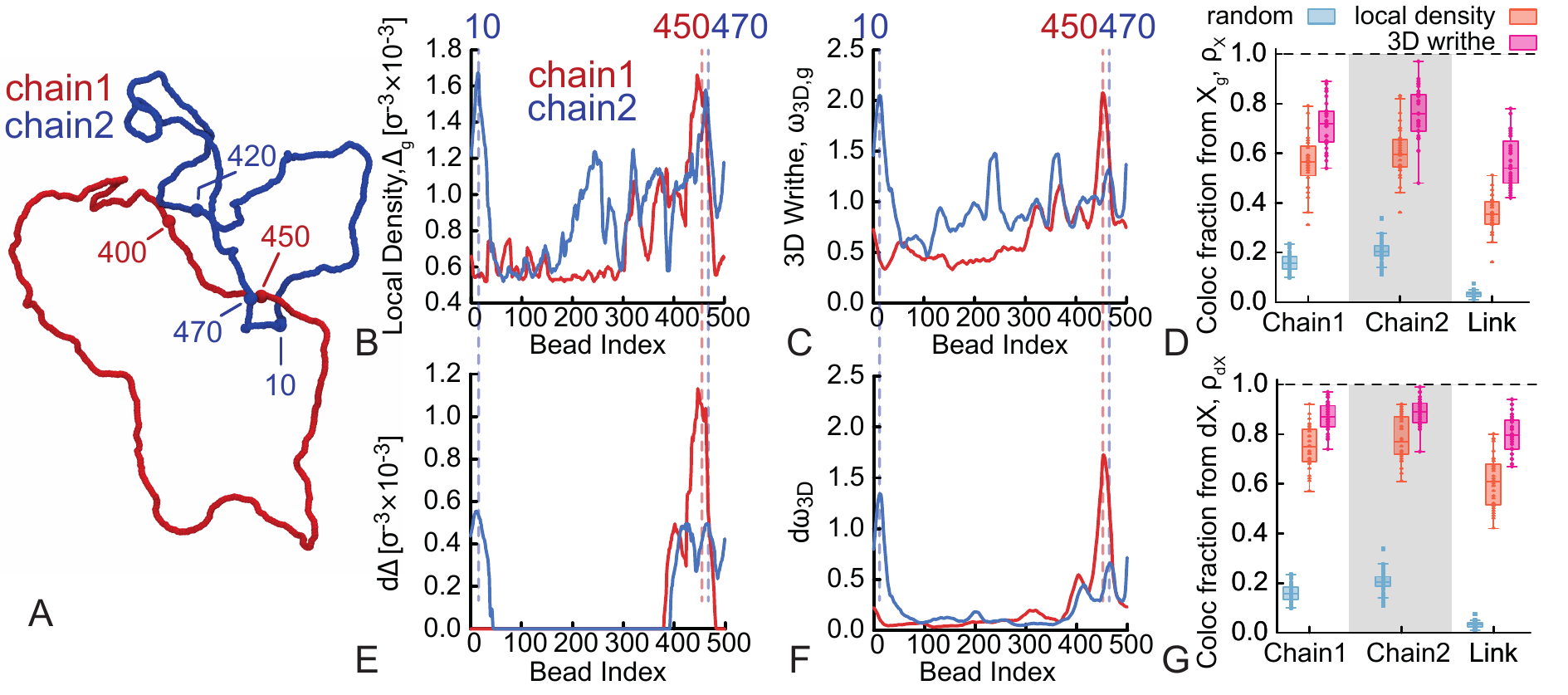}
\caption{\textbf{A.} Snapshot of a MD simulation of two rings, each $N=500$ beads long, tied in a Hopf link. Some beads are highlighted and made larger for visualisation purposes. The algorithm introduced in Ref.~\cite{Caraglio2017} detected the shortest linked segments spanning beads 421 to 460 for chain 1 and 461 to 20 for chain 2 (across the periodic boundary). \textbf{B.} Local density, $\Delta_g$, and \textbf{C.} 3D writhe, $\omega_{3D,g}$, computed considering all the beads in the system. \textbf{D.} Colocalisation score for the single chain components and the overall link from the ``global'' predictors, $X_g$. \textbf{E.} Local density, $d\Delta$, and \textbf{F.} 3D writhe, $d \omega_{3D}$, computed from the difference of global and self components of the predictors: $dX(i) = X_g(i) - X_s(i)$. \textbf{G.} Colocalisation score for the single chain components and the overall link from the differential predictors, $dX$.}
\label{fig:coloc_link}
\end{figure*}

\subsection{Link Localisation by geometric descriptors}
In the last part of this paper we consider links as prototypical examples of generic entangled chains. We perform MD simulations of two $N=500$ beads-spring Kremer-Grest polymer chains tied in a simple Hopf link. We then measure the shortest linked portion using the method described in Refs.~\cite{Caraglio2017,Caraglio2017a,Caraglio2020,Amici2019}, and compare the resulting segment with the ones given by our geometric descriptors.
Briefly, the algorithm works as follows: from a pair of linked curves with topology $\tau$ computed using the two-variable Alexander polynomial~\cite{Caraglio2017}, it is possible to obtain the shortest physical link by looking at all possible pairs of subchains ($\gamma_1$, $\gamma_2$) conditional that they display the same topology as the original link. The algorithm employs a top-down search scheme based on a bisection method and outputs the index of the beads in chain 1 and chain 2. We then count how likely it is that the $i_X$'s obtained using the geometric predictors fall within the shortest linked regions of the two chains. 

We here compare the results from the link localisation algorithm with our two best performing descriptors, i.e. the local density, $\Delta$, and the 3D writhe, $\omega_{3D}$. Since we now consider two chains, we can define $\Delta(i)$ and $\omega_{3D}(i)$ as ``self'' (when computing them considering only the chain that hosts the $i$\textsuperscript{th} segment) or as ``global'' (when considering all beads in the system in the calculation). The trend of $X_{\rm s}(i)$ reflect the entanglements of the chain with itself while $X_{\rm g}(i)$ mirrors any entanglement segment $i$ is subjected to. In Fig.~\ref{fig:coloc_link}A-C we show that for a randomly chosen simulation snapshot, the global features $X_g(i)$ display several maxima and the higher ones correspond to the beads forming the link. For the particular snapshot in Fig.~\ref{fig:coloc_link}A, the link localisation algorithm~\cite{Caraglio2017} detects the shortest linked arc in chain 1 (red in the figure) to be 421-460 and the shortest linked arc in chain 2 (blue in the figure) to be 461-20 (through periodic boundary conditions at $N=500$). We highlighted the positions of these beads in Figs.~\ref{fig:coloc_link}A-C, E and F, to show the agreement with $\Delta_g$ and $\omega_{3D,g}$.

The colocalisation score calculated on the global geometric predictors (shown in Fig.~\ref{fig:coloc_link}D) suggests that these features correlate well with the location of the link. As expected, we do not see any significant difference when comparing the accuracy of chain 1 and chain 2, and we observe that the colocalisation score for the total link, i.e. the conditional probability that both linked segments contain $i_{X_g}$, appears to be roughly the product of the two colocalisation scores for the single components. Importantly, Fig.~\ref{fig:coloc_link}D shows that the geometric predictors significantly outperform the random prediction (even by a factor of 5 or more). 

We then noted that the difference of the global and the self components of the geometric predictors, defined as $dX(i) = X_{\rm g}(i)-X_{\rm s}(i)$, significantly decrease the fluctuations of the curves. Intuitively, $dX(i)$ counts the contributions of inter-chain segments on the segment $i$ (see Fig.~\ref{fig:coloc_link}E,F). Strikingly, we find that $i_{dX}$, i.e. the bead hosting the maximum value of the difference $dX$, yields an even better colocalisation score, with values around 90\% for the individual link components and $80\%$ for the total link (Fig.~\ref{fig:coloc_link}G). The ratio of the localisation accuracy of the geometric predictors and the random choice is now 10 or more. \dmi{Arguably, this means that the  inter-chain correlations are the most important contribution to entanglements. This is also in line with the situation in entangled polymer melts, where total density fluctuations are typically small, while inter-chain density fluctuations are more informative of the system dynamics~\cite{Tsang2017,Dell2018}. }

\section{Discussion and Conclusions}
What makes a \dmi{curve} knotted? Inside our cells, how do certain proteins recognise complex topologies by scanning the DNA locally? \dmi{How can we unambiguously identify relevant entanglements in polymeric systems?} In this work we started from the hypothesis that knotted and linked curves in 3D may harbor some geometric features that correlate with the underlying topology. To this end we have performed MD simulations of knotted and linked curves and have analysed four geometric predictors: (i) local curvature, (ii) local density, (iii) 1D writhe and (iv) 3D writhe. We used the geometric predictors to locate the shortest knotted and linked arcs and compared these predictions to the ones given by state-of-the-art knot and link localisation algorithms (Refs.~\cite{Tubiana2011,Tubiana2018,Caraglio2017}). 

We discovered that local curvature is equivalent to randomly choosing a bead within the contour. This is interesting as there are models arguing that Topoisomerase, a protein involved in simplifying knots in DNA, may sense curvature to locate a knotted segment~\cite{Burnier2007}. Our work suggests that this would be a poor search strategy and would yield a rather inefficient topological simplification pathway. \dmi{Admittedly, our model does not capture the torsional rigidity and the double-helical structure of DNA and we thus refrain from arguing that our results clarify the search strategy of Topoisomerases on DNA}. \dmi{At the same time, our results suggest that in polymer melts and other generic thermally-driven entangled systems, such as weavings, the points of maximum curvature of the filaments are not necessarily the most entangled}. 

On the other hand, we find that local density is a far better geometric predictor of topologically complex states. In our simulations, the bead in the polymer with the largest number of neighbours (largest local density) is often also part of the knotted or linked segment (with accuracy $\simeq 80\%$ for simple knots and the Hopf link and up to $90\%$ for more complex knots or under confinement). This is rather striking in that the calculation of local density is restricted to beads that are 3D proximal to bead $i$ and there is no information on the global topology of the curve. One consequence of our findings is that sensing the local density of DNA segments could be a good strategy for Topoisomerase to quickly locate knotted and entangled arcs. Such a binding strategy may be naturally realised by a protein design that presents abundant positively charged amino acids on the surface of the protein, in such a way as to maximise unspecific interactions with negatively charged DNA. \dmi{Indeed, Topoisomerases typically present a positively charged area in the region of DNA-binding that is far larger than the one needed to bind DNA~\cite{Dong2007,MoraisCabral1997}.} \dmi{Again, we stress that our polymer model does not fully capture DNA's complexity. In the future we aim to perform a similar analysis on models that can capture twist~\cite{Ouldridge2013,Brackley2014} to quantify the impact of torsional rigidity on these metrics. Furthermore, it has been suggested that in knotted and closed DNA there may be an interplay of both knots and plectonemes; in this case the geometric descriptors measured here may struggle to identify the essential crossings of the knot from the writhe of the plectoneme. Future studies will illuminate this issue.}
\dmi{In spite of the limitations of our present model in modelling DNA, we conjecture that our results may be used to quantify entanglement motifs in tangled and weaved structures~\cite{August2020,Oster2021}. For instance, we expect that the pattern of local density along the entangled curves will be motif-dependent and that there may be a relationship between these patterns and the corresponding mesoscopic elasticity. Again, we hope that future work will explore this direction further.}

Finally, we discover that 3D writhe is our best descriptor with a consistently high ($\gtrsim 90\%$) accuracy in identifying the knotted and linked arcs. This observation is less striking than the one for the local density as 3D writhe is not (strictly speaking) a local geometric predictor. In other words, the calculation of 3D writhe has to scale as $N^2$ while the local curvature, density and 1D writhe scale as $N$. We note that local density can make use of neighbour lists, hence why we claim it could scale faster than $N^2$.

In line with this, we note that state-of-the-art algorithms that search for knotted and linked segments on polymeric systems~\cite{Caraglio2017,Tubiana2011closure,Tubiana2018} or proteins~\cite{Dabrowski-Tumanski2016a,Dabrowski-Tumanski2021,Dabrowski-Tumanski2017}, require a considerable amount of computational time. For instance, \dmi{when run on a single CPU}, knot localisation on our $N=500$ chain in dilute conditions takes about 2 milliseconds but under confinement takes up to 300 milliseconds per conformation. On the other hand, the calculation of the local density profile takes on average 0.3 milliseconds. Similarly, link localisation for our two N=500 chains takes up to a minute even in dilute conditions on a single conformation. On the contrary, the calculation of the local density profile for the same link takes 30 milliseconds per conformation. For this reason we argue that adding a preliminary search step using geometric predictors, before launching a full blown topological search scheme, could be a way to render search algorithms more efficient in the future. 

\dmi{It is appropriate here to highlight that entanglements are among the most elusive and slippery topics in polymer science. Algorithms such as isoconfigurational mean path~\cite{Bisbee2011} and primitive path analysis~\cite{Everaers2004} are the ``gold standard'' to quantify relevant entanglements in polymeric systems and yet they fail in the case of ring polymers~\cite{Halverson2011statics}. We hope that the geometric descriptors proposed here may be a complement to these tools and could be used to identify entanglements in complex polymeric systems. We speculate that (inter-chain) local density, 1D and 3D writhe as defined in this work may yield interesting results not only in melts of ring polymers but also in molecular (and periodic) weavings~\cite{Evans2014,Igram2016,August2020,Oster2021}. We expect that different entanglement motifs are associated with distinct patterns of our geometric observables. In turn, they may be used to predict the global elastic response of the entangled network to certain perturbations. To the best of our knowledge, these metrics have not yet been tried on polymer melts or molecular weavings.}

\dmi{One intriguing application of our results is on Olympic gels~\cite{Gennes1979,Kim2013a,Krajina2018,Klotz2020kdna}. Indeed, there is no simple way to compute the extension of three or more components of the Gauss linking number -- known as the Milnor's triple linking number~\cite{Polyak1997} -- on systems of ring polymers. This means that it is extremely challenging to unambiguously discern three, physically inseparable, Borromean rings from three unlinked, and physically separable, rings. Systems made of interlinked ``Olympic'' rings~\cite{Gennes1979}, such as the naturally occurring Kinetoplast DNA~\cite{Chen1995,Klotz2020kdna} or synthetic equivalents~\cite{Krajina2018}, are likely to display Borromean, and higher order Brunnian, configurations of interlinked rings~\cite{michieletto2015kinetoplast}. This means that computing the pair-wise (Gauss) linking number between rings is likely not enough to predict the mesoscopic elasticity of Olympic gels, as this metric completely neglects contributions from Brunnian links. We hope that our geometric predictors may be able to offer an alternative to the lack of (simple) topological invariants to characterise these elusive conformations. For instance, a step towards this goal in the near future would be to study the behaviour of our geometric predictors in simple Borromean rings in dilute conditions.  
}

Finally, we note that the data generated by our geometric predictors lend themselves fittingly to be used as input features for machine learning algorithms, e.g. neural networks, to identify knots and entanglements. This is because our predictors are invariant under translations and rotations of the conformation and under relabelling of the beads. In the future, we thus aim to couple our geometric observables to Machine Learning, as recently done in Ref.~\cite{Vandans2020}, to identify and localise knots and entanglements in more complex systems.

\begin{acknowledgement}
DM thanks the Royal Society for support through a University Research Fellowship. This project has received funding from the European Research Council (ERC) under the European Union's Horizon 2020 research and innovation programme (grant agreement No 947918, TAP). The codes to run the simulations and compute the geometric descriptors can be found open source at \url{https://git.ecdf.ed.ac.uk/taplab/geomtopo.git}. We also acknowledge insightful questions and comments from the anonymous referees.
\end{acknowledgement}


%
\bibliography{library}

\providecommand{\latin}[1]{#1}
\makeatletter
\providecommand{\doi}
  {\begingroup\let\do\@makeother\dospecials
  \catcode`\{=1 \catcode`\}=2 \doi@aux}
\providecommand{\doi@aux}[1]{\endgroup\texttt{#1}}
\makeatother
\providecommand*\mcitethebibliography{\thebibliography}
\csname @ifundefined\endcsname{endmcitethebibliography}
  {\let\endmcitethebibliography\endthebibliography}{}
\begin{mcitethebibliography}{84}
\providecommand*\natexlab[1]{#1}
\providecommand*\mciteSetBstSublistMode[1]{}
\providecommand*\mciteSetBstMaxWidthForm[2]{}
\providecommand*\mciteBstWouldAddEndPuncttrue
  {\def\EndOfBibitem{\unskip.}}
\providecommand*\mciteBstWouldAddEndPunctfalse
  {\let\EndOfBibitem\relax}
\providecommand*\mciteSetBstMidEndSepPunct[3]{}
\providecommand*\mciteSetBstSublistLabelBeginEnd[3]{}
\providecommand*\EndOfBibitem{}
\mciteSetBstSublistMode{f}
\mciteSetBstMaxWidthForm{subitem}{(\alph{mcitesubitemcount})}
\mciteSetBstSublistLabelBeginEnd
  {\mcitemaxwidthsubitemform\space}
  {\relax}
  {\relax}

\bibitem[Panagiotou \latin{et~al.}(2013)Panagiotou, Kr{\"{o}}ger, and
  Millett]{Panagiotou2013}
Panagiotou,~E.; Kr{\"{o}}ger,~M.; Millett,~K.~C. {Writhe and mutual
  entanglement combine to give the entanglement length}. \emph{Physical Review
  E - Statistical, Nonlinear, and Soft Matter Physics} \textbf{2013},
  \emph{88}, 32--35\relax
\mciteBstWouldAddEndPuncttrue
\mciteSetBstMidEndSepPunct{\mcitedefaultmidpunct}
{\mcitedefaultendpunct}{\mcitedefaultseppunct}\relax
\EndOfBibitem
\bibitem[Igram \latin{et~al.}(2016)Igram, Millett, and Panagiotou]{Igram2016}
Igram,~S.; Millett,~K.~C.; Panagiotou,~E. {Resolving critical degrees of
  entanglement in Olympic ring systems}. \emph{Journal of Knot Theory and its
  Ramifications} \textbf{2016}, \emph{25}\relax
\mciteBstWouldAddEndPuncttrue
\mciteSetBstMidEndSepPunct{\mcitedefaultmidpunct}
{\mcitedefaultendpunct}{\mcitedefaultseppunct}\relax
\EndOfBibitem
\bibitem[Tubiana \latin{et~al.}(2011)Tubiana, Orlandini, and
  Micheletti]{Tubiana2011}
Tubiana,~L.; Orlandini,~E.; Micheletti,~C. {Multiscale entanglement in ring
  polymers under spherical confinement}. \emph{Phys. Rev. Lett.} \textbf{2011},
  \emph{107}, 1--4\relax
\mciteBstWouldAddEndPuncttrue
\mciteSetBstMidEndSepPunct{\mcitedefaultmidpunct}
{\mcitedefaultendpunct}{\mcitedefaultseppunct}\relax
\EndOfBibitem
\bibitem[Goundaroulis \latin{et~al.}(2017)Goundaroulis, Dorier, Benedetti, and
  Stasiak]{Goundaroulis2017}
Goundaroulis,~D.; Dorier,~J.; Benedetti,~F.; Stasiak,~A. {Studies of global and
  local entanglements of individual protein chains using the concept of
  knotoids}. \emph{Scientific Reports} \textbf{2017}, \emph{7}, 1--9\relax
\mciteBstWouldAddEndPuncttrue
\mciteSetBstMidEndSepPunct{\mcitedefaultmidpunct}
{\mcitedefaultendpunct}{\mcitedefaultseppunct}\relax
\EndOfBibitem
\bibitem[Goundaroulis \latin{et~al.}(2017)Goundaroulis, Dorier, Benedetti, and
  Stasiak]{Goundaroulis2017a}
Goundaroulis,~D.; Dorier,~J.; Benedetti,~F.; Stasiak,~A. {Studies of global and
  local entanglements of individual protein chains using the concept of
  knotoids}. \emph{Scientific Reports} \textbf{2017}, \emph{7}, 1--9\relax
\mciteBstWouldAddEndPuncttrue
\mciteSetBstMidEndSepPunct{\mcitedefaultmidpunct}
{\mcitedefaultendpunct}{\mcitedefaultseppunct}\relax
\EndOfBibitem
\bibitem[Goundaroulis \latin{et~al.}(2020)Goundaroulis, {Lieberman Aiden}, and
  Stasiak]{Goundaroulis2019}
Goundaroulis,~D.; {Lieberman Aiden},~E.; Stasiak,~A. {Chromatin Is Frequently
  Unknotted at the Megabase Scale}. \emph{Biophysical Journal} \textbf{2020},
  \emph{118}, 2268--2279\relax
\mciteBstWouldAddEndPuncttrue
\mciteSetBstMidEndSepPunct{\mcitedefaultmidpunct}
{\mcitedefaultendpunct}{\mcitedefaultseppunct}\relax
\EndOfBibitem
\bibitem[Panagiotou and Kauffman(2020)Panagiotou, and Kauffman]{Panagiotou2020}
Panagiotou,~E.; Kauffman,~L.~H. {Knot polynomials of open and closed curves:
  Knot polynomials of open curves}. \emph{Proceedings of the Royal Society A:
  Mathematical, Physical and Engineering Sciences} \textbf{2020},
  \emph{476}\relax
\mciteBstWouldAddEndPuncttrue
\mciteSetBstMidEndSepPunct{\mcitedefaultmidpunct}
{\mcitedefaultendpunct}{\mcitedefaultseppunct}\relax
\EndOfBibitem
\bibitem[Machon and Alexander(2013)Machon, and Alexander]{Machon2013}
Machon,~T.; Alexander,~G.~P. {Knots and nonorientable surfaces in chiral
  nematics}. \emph{Proc. Natl. Acad. Sci. USA} \textbf{2013}, \emph{110},
  14174--14179\relax
\mciteBstWouldAddEndPuncttrue
\mciteSetBstMidEndSepPunct{\mcitedefaultmidpunct}
{\mcitedefaultendpunct}{\mcitedefaultseppunct}\relax
\EndOfBibitem
\bibitem[O'Holleran \latin{et~al.}(2008)O'Holleran, Dennis, Flossmann, and
  Padgett]{OHolleran2008}
O'Holleran,~K.; Dennis,~M.~R.; Flossmann,~F.; Padgett,~M.~J. {Fractality of
  light's darkness}. \emph{Physical Review Letters} \textbf{2008},
  \emph{100}\relax
\mciteBstWouldAddEndPuncttrue
\mciteSetBstMidEndSepPunct{\mcitedefaultmidpunct}
{\mcitedefaultendpunct}{\mcitedefaultseppunct}\relax
\EndOfBibitem
\bibitem[Dennis \latin{et~al.}(2010)Dennis, King, Jack, O'Holleran, and
  Padgett]{Dennis2010}
Dennis,~M.; King,~R.; Jack,~B.; O'Holleran,~K.; Padgett,~M. {Isolated optical
  vortex knots}. \emph{Nat. Phys.} \textbf{2010}, \emph{6}, 118--121\relax
\mciteBstWouldAddEndPuncttrue
\mciteSetBstMidEndSepPunct{\mcitedefaultmidpunct}
{\mcitedefaultendpunct}{\mcitedefaultseppunct}\relax
\EndOfBibitem
\bibitem[Laing \latin{et~al.}(2015)Laing, Ricca, and Sumners]{Laing2015}
Laing,~C.~E.; Ricca,~R.~L.; Sumners,~D. W.~L. {Conservation of writhe helicity
  under anti-parallel reconnection}. \emph{Sci. Rep.} \textbf{2015}, \emph{5},
  1--6\relax
\mciteBstWouldAddEndPuncttrue
\mciteSetBstMidEndSepPunct{\mcitedefaultmidpunct}
{\mcitedefaultendpunct}{\mcitedefaultseppunct}\relax
\EndOfBibitem
\bibitem[Smrek and Grosberg(2013)Smrek, and Grosberg]{Smrek2013}
Smrek,~J.; Grosberg,~A.~Y. {A novel family of space-filling curves in their
  relation to chromosome conformation in eukaryotes}. \emph{Physica A}
  \textbf{2013}, \emph{392}, 6375--6388\relax
\mciteBstWouldAddEndPuncttrue
\mciteSetBstMidEndSepPunct{\mcitedefaultmidpunct}
{\mcitedefaultendpunct}{\mcitedefaultseppunct}\relax
\EndOfBibitem
\bibitem[Siebert \latin{et~al.}(2017)Siebert, Kivel, Atkinson, Stevens, Laue,
  and Virnau]{Siebert2017}
Siebert,~J.~T.; Kivel,~A.~N.; Atkinson,~L.~P.; Stevens,~T.~J.; Laue,~E.~D.;
  Virnau,~P. {Are there knots in chromosomes?} \emph{Polymers} \textbf{2017},
  \emph{9}, 1--10\relax
\mciteBstWouldAddEndPuncttrue
\mciteSetBstMidEndSepPunct{\mcitedefaultmidpunct}
{\mcitedefaultendpunct}{\mcitedefaultseppunct}\relax
\EndOfBibitem
\bibitem[Michieletto \latin{et~al.}(2017)Michieletto, Orlandini, and
  Marenduzzo]{Michieletto2017}
Michieletto,~D.; Orlandini,~E.; Marenduzzo,~D. {Epigenetic Transitions and
  Knotted Solitons in Stretched Chromatin}. \emph{Scientific Reports}
  \textbf{2017}, \emph{7}\relax
\mciteBstWouldAddEndPuncttrue
\mciteSetBstMidEndSepPunct{\mcitedefaultmidpunct}
{\mcitedefaultendpunct}{\mcitedefaultseppunct}\relax
\EndOfBibitem
\bibitem[Marenduzzo \latin{et~al.}(2009)Marenduzzo, Orlandini, Stasiak,
  Sumners, Tubiana, and Micheletti]{Marenduzzo2009}
Marenduzzo,~D.; Orlandini,~E.; Stasiak,~A.; Sumners,~D.; Tubiana,~L.;
  Micheletti,~C. {DNA-DNA interactions in bacteriophage capsids are responsible
  for the observed DNA knotting}. \emph{Proc. Natl. Acad. Sci. USA}
  \textbf{2009}, \emph{106}, 22269--74\relax
\mciteBstWouldAddEndPuncttrue
\mciteSetBstMidEndSepPunct{\mcitedefaultmidpunct}
{\mcitedefaultendpunct}{\mcitedefaultseppunct}\relax
\EndOfBibitem
\bibitem[Klotz \latin{et~al.}(2020)Klotz, Soh, and Doyle]{Klotz2020kdna}
Klotz,~A.~R.; Soh,~B.~W.; Doyle,~P.~S. {Equilibrium structure and deformation
  response of 2D kinetoplast sheets}. \emph{Proceedings of the National Academy
  of Sciences of the United States of America} \textbf{2020}, \emph{117},
  121--127\relax
\mciteBstWouldAddEndPuncttrue
\mciteSetBstMidEndSepPunct{\mcitedefaultmidpunct}
{\mcitedefaultendpunct}{\mcitedefaultseppunct}\relax
\EndOfBibitem
\bibitem[Klotz \latin{et~al.}(2020)Klotz, Soh, and Doyle]{Klotz2020knots}
Klotz,~A.~R.; Soh,~B.~W.; Doyle,~P.~S. {An experimental investigation of
  attraction between knots in a stretched DNA molecule}. \emph{Epl}
  \textbf{2020}, \emph{129}\relax
\mciteBstWouldAddEndPuncttrue
\mciteSetBstMidEndSepPunct{\mcitedefaultmidpunct}
{\mcitedefaultendpunct}{\mcitedefaultseppunct}\relax
\EndOfBibitem
\bibitem[Polson \latin{et~al.}(2021)Polson, Garcia, and Klotz]{Polson2021}
Polson,~J.~M.; Garcia,~E.~J.; Klotz,~A.~R. {Flatness and intrinsic curvature of
  linked-ring membranes}. \emph{Soft Matter} \textbf{2021}, \emph{17},
  10505--10515\relax
\mciteBstWouldAddEndPuncttrue
\mciteSetBstMidEndSepPunct{\mcitedefaultmidpunct}
{\mcitedefaultendpunct}{\mcitedefaultseppunct}\relax
\EndOfBibitem
\bibitem[Baiesi \latin{et~al.}(2017)Baiesi, Orlandini, Seno, and
  Trovato]{Baiesi2017}
Baiesi,~M.; Orlandini,~E.; Seno,~F.; Trovato,~A. {Exploring the correlation
  between the folding rates of proteins and the entanglement of their native
  states}. \emph{Journal of Physics A: Mathematical and Theoretical}
  \textbf{2017}, \emph{50}\relax
\mciteBstWouldAddEndPuncttrue
\mciteSetBstMidEndSepPunct{\mcitedefaultmidpunct}
{\mcitedefaultendpunct}{\mcitedefaultseppunct}\relax
\EndOfBibitem
\bibitem[Dabrowski-Tumanski and Sulkowska(2017)Dabrowski-Tumanski, and
  Sulkowska]{Dabrowski-Tumanski2017}
Dabrowski-Tumanski,~P.; Sulkowska,~J.~I. {Topological knots and links in
  proteins}. \emph{Proceedings of the National Academy of Sciences}
  \textbf{2017}, \emph{114}, 3415--3420\relax
\mciteBstWouldAddEndPuncttrue
\mciteSetBstMidEndSepPunct{\mcitedefaultmidpunct}
{\mcitedefaultendpunct}{\mcitedefaultseppunct}\relax
\EndOfBibitem
\bibitem[Marenduzzo \latin{et~al.}(2010)Marenduzzo, Micheletti, and
  Orlandini]{Marenduzzo2010}
Marenduzzo,~D.; Micheletti,~C.; Orlandini,~E. {Biopolymer organization upon
  confinement.} \emph{J. Phys.: Condens. Matter} \textbf{2010}, \emph{22},
  283102\relax
\mciteBstWouldAddEndPuncttrue
\mciteSetBstMidEndSepPunct{\mcitedefaultmidpunct}
{\mcitedefaultendpunct}{\mcitedefaultseppunct}\relax
\EndOfBibitem
\bibitem[Wu \latin{et~al.}(2017)Wu, Rauscher, Lang, Wojtecki, {De Pablo}, Hore,
  and Rowan]{Wu2017}
Wu,~Q.; Rauscher,~P.~M.; Lang,~X.; Wojtecki,~R.~J.; {De Pablo},~J.~J.;
  Hore,~M.~J.; Rowan,~S.~J. {Poly[n]catenanes: Synthesis of molecular
  interlocked chains}. \emph{Science} \textbf{2017}, \emph{358},
  1434--1439\relax
\mciteBstWouldAddEndPuncttrue
\mciteSetBstMidEndSepPunct{\mcitedefaultmidpunct}
{\mcitedefaultendpunct}{\mcitedefaultseppunct}\relax
\EndOfBibitem
\bibitem[Rauscher \latin{et~al.}(2018)Rauscher, Rowan, and {De
  Pablo}]{Rauscher2018}
Rauscher,~P.~M.; Rowan,~S.~J.; {De Pablo},~J.~J. {Topological Effects in
  Isolated Poly[ n]catenanes: Molecular Dynamics Simulations and Rouse Mode
  Analysis}. \emph{ACS Macro Letters} \textbf{2018}, \emph{7}, 938--943\relax
\mciteBstWouldAddEndPuncttrue
\mciteSetBstMidEndSepPunct{\mcitedefaultmidpunct}
{\mcitedefaultendpunct}{\mcitedefaultseppunct}\relax
\EndOfBibitem
\bibitem[Goldstein \latin{et~al.}(2010)Goldstein, Moffatt, Pesci, and
  Ricca]{Goldstein2010b}
Goldstein,~R.~E.; Moffatt,~H.~K.; Pesci,~A.~I.; Ricca,~R.~L. {Soap-film Mobius
  strip changes topology with a twist singularity}. \emph{Proc. Natl. Acad.
  Sci. USA} \textbf{2010}, \emph{107}, 21979--21984\relax
\mciteBstWouldAddEndPuncttrue
\mciteSetBstMidEndSepPunct{\mcitedefaultmidpunct}
{\mcitedefaultendpunct}{\mcitedefaultseppunct}\relax
\EndOfBibitem
\bibitem[Machon \latin{et~al.}(2016)Machon, Alexander, Goldstein, and
  Pesci]{Machon2016}
Machon,~T.; Alexander,~G.~P.; Goldstein,~R.~E.; Pesci,~A.~I. {Instabilities and
  Solitons in Minimal Strips}. \textbf{2016}, \emph{017801}, 1--7\relax
\mciteBstWouldAddEndPuncttrue
\mciteSetBstMidEndSepPunct{\mcitedefaultmidpunct}
{\mcitedefaultendpunct}{\mcitedefaultseppunct}\relax
\EndOfBibitem
\bibitem[Kamien(2002)]{Kamien2002}
Kamien,~R.~D. {The geometry of soft materials: a primer}. \emph{Reviews of
  Modern Physics} \textbf{2002}, \emph{74}, 953--971\relax
\mciteBstWouldAddEndPuncttrue
\mciteSetBstMidEndSepPunct{\mcitedefaultmidpunct}
{\mcitedefaultendpunct}{\mcitedefaultseppunct}\relax
\EndOfBibitem
\bibitem[Dennis and Hannay(2005)Dennis, and Hannay]{Dennis2005}
Dennis,~M.~R.; Hannay,~J.~H. {Geometry of Cǎlugǎreanu's theorem}. \emph{Proc.
  R. Soc. A} \textbf{2005}, \emph{461}, 3245--3254\relax
\mciteBstWouldAddEndPuncttrue
\mciteSetBstMidEndSepPunct{\mcitedefaultmidpunct}
{\mcitedefaultendpunct}{\mcitedefaultseppunct}\relax
\EndOfBibitem
\bibitem[Everaers(2004)]{Everaers2004}
Everaers,~R. {Rheology and Microscopic Topology of Entangled Polymeric
  Liquids}. \emph{Science} \textbf{2004}, \emph{303}, 823--826\relax
\mciteBstWouldAddEndPuncttrue
\mciteSetBstMidEndSepPunct{\mcitedefaultmidpunct}
{\mcitedefaultendpunct}{\mcitedefaultseppunct}\relax
\EndOfBibitem
\bibitem[Bisbee \latin{et~al.}(2011)Bisbee, Qin, and Milner]{Bisbee2011}
Bisbee,~W.; Qin,~J.; Milner,~S.~T. {Finding the tube with isoconfigurational
  averaging}. \emph{Macromolecules} \textbf{2011}, \emph{44}, 8972--8980\relax
\mciteBstWouldAddEndPuncttrue
\mciteSetBstMidEndSepPunct{\mcitedefaultmidpunct}
{\mcitedefaultendpunct}{\mcitedefaultseppunct}\relax
\EndOfBibitem
\bibitem[McLeish(2002)]{McLeish2002}
McLeish,~T. {Polymers without beginning or end.} \emph{Science} \textbf{2002},
  \emph{297}, 2005--6\relax
\mciteBstWouldAddEndPuncttrue
\mciteSetBstMidEndSepPunct{\mcitedefaultmidpunct}
{\mcitedefaultendpunct}{\mcitedefaultseppunct}\relax
\EndOfBibitem
\bibitem[Halverson \latin{et~al.}(2011)Halverson, Lee, Grest, Grosberg, and
  Kremer]{Halverson2011statics}
Halverson,~J.~D.; Lee,~W.~B.; Grest,~G.~S.; Grosberg,~A.~Y.; Kremer,~K.
  {Molecular dynamics simulation study of nonconcatenated ring polymers in a
  melt. I. Statics.} \emph{The Journal of chemical physics} \textbf{2011},
  \emph{134}, 204904\relax
\mciteBstWouldAddEndPuncttrue
\mciteSetBstMidEndSepPunct{\mcitedefaultmidpunct}
{\mcitedefaultendpunct}{\mcitedefaultseppunct}\relax
\EndOfBibitem
\bibitem[Rosa and Everaers(2014)Rosa, and Everaers]{Rosa2013}
Rosa,~A.; Everaers,~R. {Ring polymers in the melt state: the physics of
  crumpling}. \emph{Phys. Rev. Lett.} \textbf{2014}, \emph{112}, 118302\relax
\mciteBstWouldAddEndPuncttrue
\mciteSetBstMidEndSepPunct{\mcitedefaultmidpunct}
{\mcitedefaultendpunct}{\mcitedefaultseppunct}\relax
\EndOfBibitem
\bibitem[Halverson \latin{et~al.}(2011)Halverson, Lee, Grest, Grosberg, and
  Kremer]{Halverson2011dynamics}
Halverson,~J.~D.; Lee,~W.~B.; Grest,~G.~S.; Grosberg,~A.~Y.; Kremer,~K.
  {Molecular dynamics simulation study of nonconcatenated ring polymers in a
  melt. II. Dynamics}. \emph{The Journal of chemical physics} \textbf{2011},
  \emph{134}, 204905\relax
\mciteBstWouldAddEndPuncttrue
\mciteSetBstMidEndSepPunct{\mcitedefaultmidpunct}
{\mcitedefaultendpunct}{\mcitedefaultseppunct}\relax
\EndOfBibitem
\bibitem[Michieletto \latin{et~al.}(2014)Michieletto, Marenduzzo, Orlandini,
  Alexander, and Turner]{michieletto2014threading}
Michieletto,~D.; Marenduzzo,~D.; Orlandini,~E.; Alexander,~G.~P.; Turner,~M.~S.
  {Threading Dynamics of Ring Polymers in a Gel}. \emph{ACS MacroLetters}
  \textbf{2014}, \emph{3}, 255--259\relax
\mciteBstWouldAddEndPuncttrue
\mciteSetBstMidEndSepPunct{\mcitedefaultmidpunct}
{\mcitedefaultendpunct}{\mcitedefaultseppunct}\relax
\EndOfBibitem
\bibitem[G{\'{o}}mez \latin{et~al.}(2020)G{\'{o}}mez, Garc{\'{i}}a, and
  P{\"{o}}schel]{Gomez2020}
G{\'{o}}mez,~L.~R.; Garc{\'{i}}a,~N.~A.; P{\"{o}}schel,~T. {Packing structure
  of semiflexible rings}. \emph{Proceedings of the National Academy of Sciences
  of the United States of America} \textbf{2020}, \emph{117}, 3382--3387\relax
\mciteBstWouldAddEndPuncttrue
\mciteSetBstMidEndSepPunct{\mcitedefaultmidpunct}
{\mcitedefaultendpunct}{\mcitedefaultseppunct}\relax
\EndOfBibitem
\bibitem[Ge \latin{et~al.}(2016)Ge, Panyukov, and Rubinstein]{Ge2016}
Ge,~T.; Panyukov,~S.; Rubinstein,~M. {Self-Similar Conformations and Dynamics
  in Entangled Melts and Solutions of Nonconcatenated Ring Polymers}.
  \emph{Macromolecules} \textbf{2016}, \emph{49}, 708--722\relax
\mciteBstWouldAddEndPuncttrue
\mciteSetBstMidEndSepPunct{\mcitedefaultmidpunct}
{\mcitedefaultendpunct}{\mcitedefaultseppunct}\relax
\EndOfBibitem
\bibitem[Michieletto and Sakaue(2020)Michieletto, and
  Sakaue]{michieletto2020dynamical}
Michieletto,~D.; Sakaue,~T. {Dynamical Entanglement and Cooperative Dynamics in
  Entangled Solutions of Ring and Linear Polymers}. \emph{ACS Macro Letters}
  \textbf{2020}, \emph{10}, 129--134\relax
\mciteBstWouldAddEndPuncttrue
\mciteSetBstMidEndSepPunct{\mcitedefaultmidpunct}
{\mcitedefaultendpunct}{\mcitedefaultseppunct}\relax
\EndOfBibitem
\bibitem[Smrek and Grosberg(2016)Smrek, and Grosberg]{Smrek2016}
Smrek,~J.; Grosberg,~A.~Y. {Minimal Surfaces on Unconcatenated Polymer Rings in
  Melt}. \emph{ACS Macro Lett.} \textbf{2016}, \emph{5}, 750--754\relax
\mciteBstWouldAddEndPuncttrue
\mciteSetBstMidEndSepPunct{\mcitedefaultmidpunct}
{\mcitedefaultendpunct}{\mcitedefaultseppunct}\relax
\EndOfBibitem
\bibitem[Landuzzi \latin{et~al.}(2020)Landuzzi, Nakamura, Michieletto, and
  Sakaue]{landuzzi2020persistence}
Landuzzi,~F.; Nakamura,~T.; Michieletto,~D.; Sakaue,~T. {Persistence homology
  of entangled rings}. \emph{Physical Review Research} \textbf{2020}, \emph{2},
  33529\relax
\mciteBstWouldAddEndPuncttrue
\mciteSetBstMidEndSepPunct{\mcitedefaultmidpunct}
{\mcitedefaultendpunct}{\mcitedefaultseppunct}\relax
\EndOfBibitem
\bibitem[Evans and Roth(2014)Evans, and Roth]{Evans2014}
Evans,~M.~E.; Roth,~R. {Shaping the skin: The interplay of mesoscale geometry
  and corneocyte swelling}. \emph{Physical Review Letters} \textbf{2014},
  \emph{112}, 1--5\relax
\mciteBstWouldAddEndPuncttrue
\mciteSetBstMidEndSepPunct{\mcitedefaultmidpunct}
{\mcitedefaultendpunct}{\mcitedefaultseppunct}\relax
\EndOfBibitem
\bibitem[Oster \latin{et~al.}(2021)Oster, Dias, de~Wolff, and Evans]{Oster2021}
Oster,~M.; Dias,~M.~A.; de~Wolff,~T.; Evans,~M.~E. {Reentrant tensegrity: A
  three-periodic, chiral, tensegrity structure that is auxetic}. \emph{Science
  Advances} \textbf{2021}, \emph{7}, 1--7\relax
\mciteBstWouldAddEndPuncttrue
\mciteSetBstMidEndSepPunct{\mcitedefaultmidpunct}
{\mcitedefaultendpunct}{\mcitedefaultseppunct}\relax
\EndOfBibitem
\bibitem[August \latin{et~al.}(2020)August, Dryfe, Haigh, Kent, Leigh,
  Lemonnier, Li, Muryn, Palmer, Song, Whitehead, and Young]{August2020}
August,~D.~P.; Dryfe,~R.~A.; Haigh,~S.~J.; Kent,~P.~R.; Leigh,~D.~A.;
  Lemonnier,~J.~F.; Li,~Z.; Muryn,~C.~A.; Palmer,~L.~I.; Song,~Y.;
  Whitehead,~G.~F.; Young,~R.~J. {Self-assembly of a layered two-dimensional
  molecularly woven fabric}. \emph{Nature} \textbf{2020}, \emph{588},
  429--435\relax
\mciteBstWouldAddEndPuncttrue
\mciteSetBstMidEndSepPunct{\mcitedefaultmidpunct}
{\mcitedefaultendpunct}{\mcitedefaultseppunct}\relax
\EndOfBibitem
\bibitem[Matsumoto \latin{et~al.}(2018)Matsumoto, Liang, and
  Mahadevan]{Matsumoto2018}
Matsumoto,~E.~A.; Liang,~H.; Mahadevan,~L. {Topology, Geometry, and Mechanics
  of Z -Plasty}. \emph{Physical Review Letters} \textbf{2018}, \emph{120},
  68101\relax
\mciteBstWouldAddEndPuncttrue
\mciteSetBstMidEndSepPunct{\mcitedefaultmidpunct}
{\mcitedefaultendpunct}{\mcitedefaultseppunct}\relax
\EndOfBibitem
\bibitem[Markande and Matsumoto(2020)Markande, and Matsumoto]{Markande2020}
Markande,~S.~G.; Matsumoto,~E.~A. {Knotty knits are tangles on tori}.
  \emph{arxiv:2002.01497} \textbf{2020}, \relax
\mciteBstWouldAddEndPunctfalse
\mciteSetBstMidEndSepPunct{\mcitedefaultmidpunct}
{}{\mcitedefaultseppunct}\relax
\EndOfBibitem
\bibitem[Wang(1985)]{Wang1985}
Wang,~J.~C. {DNA topoisomerases.} \emph{Annu. Rev. Biochem.} \textbf{1985},
  \emph{54}, 665--97\relax
\mciteBstWouldAddEndPuncttrue
\mciteSetBstMidEndSepPunct{\mcitedefaultmidpunct}
{\mcitedefaultendpunct}{\mcitedefaultseppunct}\relax
\EndOfBibitem
\bibitem[Mart{\'{i}}nez-Garc{\'{i}}a
  \latin{et~al.}(2014)Mart{\'{i}}nez-Garc{\'{i}}a, Fern{\'{a}}ndez,
  D{\'{i}}az-Ingelmo, Rodr{\'{i}}guez-Campos, Manichanh, and
  Roca]{Martinez-Garcia2014}
Mart{\'{i}}nez-Garc{\'{i}}a,~B.; Fern{\'{a}}ndez,~X.; D{\'{i}}az-Ingelmo,~O.;
  Rodr{\'{i}}guez-Campos,~A.; Manichanh,~C.; Roca,~J. {Topoisomerase II
  minimizes DNA entanglements by proofreading DNA topology after DNA strand
  passage}. \emph{Nucleic Acids Research} \textbf{2014}, \emph{42},
  1821--1830\relax
\mciteBstWouldAddEndPuncttrue
\mciteSetBstMidEndSepPunct{\mcitedefaultmidpunct}
{\mcitedefaultendpunct}{\mcitedefaultseppunct}\relax
\EndOfBibitem
\bibitem[Michieletto \latin{et~al.}(2022)Michieletto, Fosado, Melas, Baiesi,
  Tubiana, and Orlandini]{Michieletto2022nar}
Michieletto,~D.; Fosado,~Y. A.~G.; Melas,~E.; Baiesi,~M.; Tubiana,~L.;
  Orlandini,~E. {Dynamic and facilitated binding of topoisomerase accelerates
  topological relaxation}. \textbf{2022}, \emph{50}, 4659--4668\relax
\mciteBstWouldAddEndPuncttrue
\mciteSetBstMidEndSepPunct{\mcitedefaultmidpunct}
{\mcitedefaultendpunct}{\mcitedefaultseppunct}\relax
\EndOfBibitem
\bibitem[Piskadlo and Oliveira(2017)Piskadlo, and Oliveira]{Piskadlo2017}
Piskadlo,~E.; Oliveira,~R.~A. {A topology-centric view on mitotic chromosome
  architecture}. \emph{International Journal of Molecular Sciences}
  \textbf{2017}, \emph{18}, 1--21\relax
\mciteBstWouldAddEndPuncttrue
\mciteSetBstMidEndSepPunct{\mcitedefaultmidpunct}
{\mcitedefaultendpunct}{\mcitedefaultseppunct}\relax
\EndOfBibitem
\bibitem[Vald{\'{e}}s \latin{et~al.}(2018)Vald{\'{e}}s, Segura, Dyson,
  Mart{\'{i}}nez-Garc{\'{i}}a, and Roca]{Valdes2018}
Vald{\'{e}}s,~A.; Segura,~J.; Dyson,~S.; Mart{\'{i}}nez-Garc{\'{i}}a,~B.;
  Roca,~J. {DNA knots occur in intracellular chromatin}. \emph{Nucleic Acids
  Research} \textbf{2018}, \emph{46}, 650--660\relax
\mciteBstWouldAddEndPuncttrue
\mciteSetBstMidEndSepPunct{\mcitedefaultmidpunct}
{\mcitedefaultendpunct}{\mcitedefaultseppunct}\relax
\EndOfBibitem
\bibitem[Rybenkov \latin{et~al.}(1997)Rybenkov, Ullsperger, Vologodskii,
  Nicholas, and Cozzarelli]{Rybenkov1997}
Rybenkov,~V.~V.; Ullsperger,~C.; Vologodskii,~A.~V.; Nicholas,~R.;
  Cozzarelli,~N.~R. {Simplification of DNA Topology Below Equilibrium Values by
  by Type II Topoisomerases}. \emph{Science} \textbf{1997}, \emph{277},
  690--693\relax
\mciteBstWouldAddEndPuncttrue
\mciteSetBstMidEndSepPunct{\mcitedefaultmidpunct}
{\mcitedefaultendpunct}{\mcitedefaultseppunct}\relax
\EndOfBibitem
\bibitem[Kremer and Grest(1990)Kremer, and Grest]{Kremer1990}
Kremer,~K.; Grest,~G.~S. {Dynamics of entangled linear polymer melts: A
  molecular-dynamics simulation}. \emph{The Journal of Chemical Physics}
  \textbf{1990}, \emph{92}, 5057--5086\relax
\mciteBstWouldAddEndPuncttrue
\mciteSetBstMidEndSepPunct{\mcitedefaultmidpunct}
{\mcitedefaultendpunct}{\mcitedefaultseppunct}\relax
\EndOfBibitem
\bibitem[Calladine \latin{et~al.}(1997)Calladine, Drew, Luisi, Travers, and
  Bash]{Calladine1997}
Calladine,~C.~R.; Drew,~H.; Luisi,~F.~B.; Travers,~A.~A.; Bash,~E.
  \emph{{Understanding DNA: the molecule and how it works}}; Elsevier Academic
  Press, 1997; Vol.~1\relax
\mciteBstWouldAddEndPuncttrue
\mciteSetBstMidEndSepPunct{\mcitedefaultmidpunct}
{\mcitedefaultendpunct}{\mcitedefaultseppunct}\relax
\EndOfBibitem
\bibitem[Plimpton(1995)]{Plimpton1995}
Plimpton,~S. {Fast Parallel Algorithms for Short-Range Molecular Dynamics}.
  \emph{J. Comp. Phys.} \textbf{1995}, \emph{117}, 1--19\relax
\mciteBstWouldAddEndPuncttrue
\mciteSetBstMidEndSepPunct{\mcitedefaultmidpunct}
{\mcitedefaultendpunct}{\mcitedefaultseppunct}\relax
\EndOfBibitem
\bibitem[Stasiak \latin{et~al.}(1996)Stasiak, Katritch, Bednar, Michoud, and
  Dubochet]{Stasiak1996}
Stasiak,~A.; Katritch,~V.; Bednar,~J.; Michoud,~D.; Dubochet,~J.
  {Electrophoretic mobility of DNA knots}. \emph{Nature} \textbf{1996},
  \emph{384}, 122\relax
\mciteBstWouldAddEndPuncttrue
\mciteSetBstMidEndSepPunct{\mcitedefaultmidpunct}
{\mcitedefaultendpunct}{\mcitedefaultseppunct}\relax
\EndOfBibitem
\bibitem[Klenin and Langowski(2000)Klenin, and Langowski]{Klenin2000}
Klenin,~K.; Langowski,~J. {Computation of writhe in modeling of supercoiled
  DNA.} \emph{Biopolymers} \textbf{2000}, \emph{54}, 307--17\relax
\mciteBstWouldAddEndPuncttrue
\mciteSetBstMidEndSepPunct{\mcitedefaultmidpunct}
{\mcitedefaultendpunct}{\mcitedefaultseppunct}\relax
\EndOfBibitem
\bibitem[Smrek \latin{et~al.}(2021)Smrek, Garamella, Robertson-Anderson, and
  Michieletto]{Smrek2021supercoiling}
Smrek,~J.; Garamella,~J.; Robertson-Anderson,~R.; Michieletto,~D. {Topological
  tuning of DNA mobility in entangled solutions of supercoiled plasmids}.
  \emph{Science Advances} \textbf{2021}, \emph{7}, 1--28\relax
\mciteBstWouldAddEndPuncttrue
\mciteSetBstMidEndSepPunct{\mcitedefaultmidpunct}
{\mcitedefaultendpunct}{\mcitedefaultseppunct}\relax
\EndOfBibitem
\bibitem[Michieletto(2016)]{Michieletto2016a}
Michieletto,~D. {On the tree-like structure of rings in dense solutions}.
  \emph{Soft Matter} \textbf{2016}, \emph{12}, 9485--9500\relax
\mciteBstWouldAddEndPuncttrue
\mciteSetBstMidEndSepPunct{\mcitedefaultmidpunct}
{\mcitedefaultendpunct}{\mcitedefaultseppunct}\relax
\EndOfBibitem
\bibitem[Tubiana \latin{et~al.}(2018)Tubiana, Polles, Orlandini, and
  Micheletti]{Tubiana2018}
Tubiana,~L.; Polles,~G.; Orlandini,~E.; Micheletti,~C. {KymoKnot: A web server
  and software package to identify and locate knots in trajectories of linear
  or circular polymers}. \emph{European Physical Journal E} \textbf{2018},
  \emph{41}\relax
\mciteBstWouldAddEndPuncttrue
\mciteSetBstMidEndSepPunct{\mcitedefaultmidpunct}
{\mcitedefaultendpunct}{\mcitedefaultseppunct}\relax
\EndOfBibitem
\bibitem[Suma and Micheletti(2017)Suma, and Micheletti]{Suma2017}
Suma,~A.; Micheletti,~C. {Pore translocation of knotted DNA rings}.
  \emph{Proceedings of the National Academy of Sciences of the United States of
  America} \textbf{2017}, \emph{114}, E2991--E2997\relax
\mciteBstWouldAddEndPuncttrue
\mciteSetBstMidEndSepPunct{\mcitedefaultmidpunct}
{\mcitedefaultendpunct}{\mcitedefaultseppunct}\relax
\EndOfBibitem
\bibitem[Coronel \latin{et~al.}(2018)Coronel, Suma, and
  Micheletti]{Coronel2018}
Coronel,~L.; Suma,~A.; Micheletti,~C. {Dynamics of supercoiled DNA with complex
  knots: Large-scale rearrangements and persistent multi-strand interlocking}.
  \emph{Nucleic Acids Research} \textbf{2018}, \emph{46}, 7533--7541\relax
\mciteBstWouldAddEndPuncttrue
\mciteSetBstMidEndSepPunct{\mcitedefaultmidpunct}
{\mcitedefaultendpunct}{\mcitedefaultseppunct}\relax
\EndOfBibitem
\bibitem[Grosberg and Rabin(2007)Grosberg, and Rabin]{Grosberg2007}
Grosberg,~A.~Y.; Rabin,~Y. {Metastable tight knots in a wormlike polymer}.
  \emph{Physical Review Letters} \textbf{2007}, \emph{99}, 1--4\relax
\mciteBstWouldAddEndPuncttrue
\mciteSetBstMidEndSepPunct{\mcitedefaultmidpunct}
{\mcitedefaultendpunct}{\mcitedefaultseppunct}\relax
\EndOfBibitem
\bibitem[Tubiana and Rosa(2013)Tubiana, and Rosa]{Tubiana2013}
Tubiana,~L.; Rosa,~A. {Spontaneous knotting and unknotting of flexible linear
  polymers: equilibrium and kinetic aspects}. \emph{arXiv pr@eprint arXiv:
  {\ldots}} \textbf{2013}, \emph{2}, 1--14\relax
\mciteBstWouldAddEndPuncttrue
\mciteSetBstMidEndSepPunct{\mcitedefaultmidpunct}
{\mcitedefaultendpunct}{\mcitedefaultseppunct}\relax
\EndOfBibitem
\bibitem[Caraglio \latin{et~al.}(2017)Caraglio, Micheletti, and
  Orlandini]{Caraglio2017}
Caraglio,~M.; Micheletti,~C.; Orlandini,~E. {Physical Links: Defining and
  detecting inter-chain entanglement}. \emph{Scientific Reports} \textbf{2017},
  \emph{7}, 1--10\relax
\mciteBstWouldAddEndPuncttrue
\mciteSetBstMidEndSepPunct{\mcitedefaultmidpunct}
{\mcitedefaultendpunct}{\mcitedefaultseppunct}\relax
\EndOfBibitem
\bibitem[Caraglio \latin{et~al.}(2017)Caraglio, Micheletti, and
  Orlandini]{Caraglio2017a}
Caraglio,~M.; Micheletti,~C.; Orlandini,~E. {Mechanical pulling of linked ring
  polymers: Elastic response and link localisation}. \emph{Polymers}
  \textbf{2017}, \emph{9}, 1--12\relax
\mciteBstWouldAddEndPuncttrue
\mciteSetBstMidEndSepPunct{\mcitedefaultmidpunct}
{\mcitedefaultendpunct}{\mcitedefaultseppunct}\relax
\EndOfBibitem
\bibitem[Caraglio \latin{et~al.}(2020)Caraglio, Orlandini, and
  Whittington]{Caraglio2020}
Caraglio,~M.; Orlandini,~E.; Whittington,~S.~G. {Translocation of links through
  a pore: Effects of link complexity and size}. \emph{Journal of Statistical
  Mechanics: Theory and Experiment} \textbf{2020}, \emph{2020}\relax
\mciteBstWouldAddEndPuncttrue
\mciteSetBstMidEndSepPunct{\mcitedefaultmidpunct}
{\mcitedefaultendpunct}{\mcitedefaultseppunct}\relax
\EndOfBibitem
\bibitem[Amici \latin{et~al.}(2019)Amici, Caraglio, Orlandini, and
  Micheletti]{Amici2019}
Amici,~G.; Caraglio,~M.; Orlandini,~E.; Micheletti,~C. {Topologically Linked
  Chains in Confinement}. \emph{ACS Macro Letters} \textbf{2019}, \emph{8},
  442--446\relax
\mciteBstWouldAddEndPuncttrue
\mciteSetBstMidEndSepPunct{\mcitedefaultmidpunct}
{\mcitedefaultendpunct}{\mcitedefaultseppunct}\relax
\EndOfBibitem
\bibitem[Tsang \latin{et~al.}(2017)Tsang, Dell, Jiang, Schweizer, and
  Granick]{Tsang2017}
Tsang,~B.; Dell,~Z.~E.; Jiang,~L.; Schweizer,~K.~S.; Granick,~S. {Dynamic
  cross-correlations between entangled biofilaments as they diffuse}.
  \emph{Proceedings of the National Academy of Sciences of the United States of
  America} \textbf{2017}, \emph{114}, 3322--3327\relax
\mciteBstWouldAddEndPuncttrue
\mciteSetBstMidEndSepPunct{\mcitedefaultmidpunct}
{\mcitedefaultendpunct}{\mcitedefaultseppunct}\relax
\EndOfBibitem
\bibitem[Dell and Schweizer(2018)Dell, and Schweizer]{Dell2018}
Dell,~Z.~E.; Schweizer,~K.~S. {Intermolecular structural correlations in model
  globular and unconcatenated ring polymer liquids}. \emph{Soft Matter}
  \textbf{2018}, \emph{14}, 9132--9142\relax
\mciteBstWouldAddEndPuncttrue
\mciteSetBstMidEndSepPunct{\mcitedefaultmidpunct}
{\mcitedefaultendpunct}{\mcitedefaultseppunct}\relax
\EndOfBibitem
\bibitem[Burnier \latin{et~al.}(2007)Burnier, Weber, Flammini, and
  Stasiak]{Burnier2007}
Burnier,~Y.; Weber,~C.; Flammini,~A.; Stasiak,~A. {Local selection rules that
  can determine specific pathways of DNA unknotting by type II DNA
  topoisomerases}. \emph{Nucleic Acids Research} \textbf{2007}, \emph{35},
  5223--5231\relax
\mciteBstWouldAddEndPuncttrue
\mciteSetBstMidEndSepPunct{\mcitedefaultmidpunct}
{\mcitedefaultendpunct}{\mcitedefaultseppunct}\relax
\EndOfBibitem
\bibitem[Dong and Berger(2007)Dong, and Berger]{Dong2007}
Dong,~K.~C.; Berger,~J.~M. {Structural basis for gate-DNA recognition and
  bending by type IIA topoisomerases.} \emph{Nature} \textbf{2007}, \emph{450},
  1201--1205\relax
\mciteBstWouldAddEndPuncttrue
\mciteSetBstMidEndSepPunct{\mcitedefaultmidpunct}
{\mcitedefaultendpunct}{\mcitedefaultseppunct}\relax
\EndOfBibitem
\bibitem[{Morais Cabral} \latin{et~al.}(1997){Morais Cabral}, Jackson, Smith,
  Shikotra, Maxwell, and Liddington]{MoraisCabral1997}
{Morais Cabral},~J.~H.; Jackson,~A.~P.; Smith,~C.~V.; Shikotra,~N.;
  Maxwell,~A.; Liddington,~R.~C. {Crystal structure of the breakage-reunion
  domain of DNA gyrase}. \emph{Nature} \textbf{1997}, \emph{388},
  903--906\relax
\mciteBstWouldAddEndPuncttrue
\mciteSetBstMidEndSepPunct{\mcitedefaultmidpunct}
{\mcitedefaultendpunct}{\mcitedefaultseppunct}\relax
\EndOfBibitem
\bibitem[Ouldridge \latin{et~al.}(2013)Ouldridge, Hoare, Louis, Doye, Bath, and
  Turberfield]{Ouldridge2013}
Ouldridge,~T.~E.; Hoare,~R.~L.; Louis,~A.~a.; Doye,~J. P.~K.; Bath,~J.;
  Turberfield,~A.~J. {Optimizing DNA nanotechnology through coarse-grained
  modeling: A two-footed DNA walker}. \emph{ACS Nano} \textbf{2013}, \emph{7},
  2479--2490\relax
\mciteBstWouldAddEndPuncttrue
\mciteSetBstMidEndSepPunct{\mcitedefaultmidpunct}
{\mcitedefaultendpunct}{\mcitedefaultseppunct}\relax
\EndOfBibitem
\bibitem[Brackley \latin{et~al.}(2014)Brackley, Morozov, and
  Marenduzzo]{Brackley2014}
Brackley,~C.~A.; Morozov,~A.~N.; Marenduzzo,~D. {Models for twistable elastic
  polymers in Brownian dynamics, and their implementation for LAMMPS.} \emph{J.
  Chem. Phys.} \textbf{2014}, \emph{140}, 135103\relax
\mciteBstWouldAddEndPuncttrue
\mciteSetBstMidEndSepPunct{\mcitedefaultmidpunct}
{\mcitedefaultendpunct}{\mcitedefaultseppunct}\relax
\EndOfBibitem
\bibitem[Tubiana \latin{et~al.}(2011)Tubiana, Orlandini, and
  Micheletti]{Tubiana2011closure}
Tubiana,~L.; Orlandini,~E.; Micheletti,~C. {Probing the Entanglement and
  Locating Knots in Ring Polymers: A Comparative Study of Different Arc Closure
  Schemes}. \emph{Prog. Theor. Phys. Suppl.} \textbf{2011}, \emph{191},
  192--204\relax
\mciteBstWouldAddEndPuncttrue
\mciteSetBstMidEndSepPunct{\mcitedefaultmidpunct}
{\mcitedefaultendpunct}{\mcitedefaultseppunct}\relax
\EndOfBibitem
\bibitem[Dabrowski-Tumanski \latin{et~al.}(2016)Dabrowski-Tumanski, Niemyska,
  Pasznik, and Sulkowska]{Dabrowski-Tumanski2016a}
Dabrowski-Tumanski,~P.; Niemyska,~W.; Pasznik,~P.; Sulkowska,~J.~I. {LassoProt:
  server to analyze biopolymers with lassos}. \emph{Nucleic acids research}
  \textbf{2016}, \emph{44}, W383--W389\relax
\mciteBstWouldAddEndPuncttrue
\mciteSetBstMidEndSepPunct{\mcitedefaultmidpunct}
{\mcitedefaultendpunct}{\mcitedefaultseppunct}\relax
\EndOfBibitem
\bibitem[Dabrowski-Tumanski \latin{et~al.}(2021)Dabrowski-Tumanski, Rubach,
  Niemyska, Gren, and Sulkowska]{Dabrowski-Tumanski2021}
Dabrowski-Tumanski,~P.; Rubach,~P.; Niemyska,~W.; Gren,~B.~A.; Sulkowska,~J.~I.
  {Topoly: Python package to analyze topology of polymers}. \emph{Briefings in
  Bioinformatics} \textbf{2021}, \emph{22}, 1--8\relax
\mciteBstWouldAddEndPuncttrue
\mciteSetBstMidEndSepPunct{\mcitedefaultmidpunct}
{\mcitedefaultendpunct}{\mcitedefaultseppunct}\relax
\EndOfBibitem
\bibitem[Gennes(1979)]{Gennes1979}
Gennes,~P. G.~D. \emph{{Scaling concepts in polymer physics}}; 1979\relax
\mciteBstWouldAddEndPuncttrue
\mciteSetBstMidEndSepPunct{\mcitedefaultmidpunct}
{\mcitedefaultendpunct}{\mcitedefaultseppunct}\relax
\EndOfBibitem
\bibitem[Kim \latin{et~al.}(2013)Kim, Kundukad, Allahverdi, Nordensk{\"{o}}ld,
  Doyle, and {Van Der Maarel}]{Kim2013a}
Kim,~Y.~S.; Kundukad,~B.; Allahverdi,~A.; Nordensk{\"{o}}ld,~L.; Doyle,~P.~S.;
  {Van Der Maarel},~J.~R. {Gelation of the genome by topoisomerase II targeting
  anticancer agents}. \emph{Soft Matter} \textbf{2013}, \emph{9},
  1656--1663\relax
\mciteBstWouldAddEndPuncttrue
\mciteSetBstMidEndSepPunct{\mcitedefaultmidpunct}
{\mcitedefaultendpunct}{\mcitedefaultseppunct}\relax
\EndOfBibitem
\bibitem[Krajina \latin{et~al.}(2018)Krajina, Zhu, Heilshorn, and
  Spakowitz]{Krajina2018}
Krajina,~B.~A.; Zhu,~A.; Heilshorn,~S.~C.; Spakowitz,~A.~J. {Active DNA Olympic
  Hydrogels Driven by Topoisomerase Activity}. \emph{Physical Review Letters}
  \textbf{2018}, \emph{121}, 148001\relax
\mciteBstWouldAddEndPuncttrue
\mciteSetBstMidEndSepPunct{\mcitedefaultmidpunct}
{\mcitedefaultendpunct}{\mcitedefaultseppunct}\relax
\EndOfBibitem
\bibitem[Polyak(1997)]{Polyak1997}
Polyak,~M. {On Milnor 's triple linking number}. \emph{C. R. Acad. Sc. Paris}
  \textbf{1997}, \emph{325}, 77--82\relax
\mciteBstWouldAddEndPuncttrue
\mciteSetBstMidEndSepPunct{\mcitedefaultmidpunct}
{\mcitedefaultendpunct}{\mcitedefaultseppunct}\relax
\EndOfBibitem
\bibitem[Chen \latin{et~al.}(1995)Chen, Rauch, White, Englund, and
  Cozzarelli]{Chen1995}
Chen,~J.; Rauch,~C.~A.; White,~J.~H.; Englund,~P.~T.; Cozzarelli,~N. {The
  topology of the kinetoplast DNA network.} \emph{Cell} \textbf{1995},
  \emph{80}, 61--9\relax
\mciteBstWouldAddEndPuncttrue
\mciteSetBstMidEndSepPunct{\mcitedefaultmidpunct}
{\mcitedefaultendpunct}{\mcitedefaultseppunct}\relax
\EndOfBibitem
\bibitem[Michieletto \latin{et~al.}(2015)Michieletto, Marenduzzo, and
  Orlandini]{michieletto2015kinetoplast}
Michieletto,~D.; Marenduzzo,~D.; Orlandini,~E. {Is the Kinetoplast DNA a
  Percolating Network of Linked Rings at its Critical Point?} \emph{Physical
  Biology} \textbf{2015}, \emph{12}, 36001\relax
\mciteBstWouldAddEndPuncttrue
\mciteSetBstMidEndSepPunct{\mcitedefaultmidpunct}
{\mcitedefaultendpunct}{\mcitedefaultseppunct}\relax
\EndOfBibitem
\bibitem[Vandans \latin{et~al.}(2020)Vandans, Yang, Wu, and Dai]{Vandans2020}
Vandans,~O.; Yang,~K.; Wu,~Z.; Dai,~L. {Identifying knot types of polymer
  conformations by machine learning}. \emph{Physical Review E} \textbf{2020},
  \emph{101}, 1--10\relax
\mciteBstWouldAddEndPuncttrue
\mciteSetBstMidEndSepPunct{\mcitedefaultmidpunct}
{\mcitedefaultendpunct}{\mcitedefaultseppunct}\relax
\EndOfBibitem
\end{mcitethebibliography}

\end{document}